\RequirePackage{etoolbox}
\csdef{input@path}{%
}%

\documentclass[ba]{imsart}
\usepackage{latexsym,amssymb,amsmath,amsfonts,color,fancyvrb,enumerate,natbib,bm,booktabs}
\usepackage{mathrsfs}

\usepackage{amsmath}
\usepackage{graphicx, psfrag, epsf}
\usepackage{enumerate}
\usepackage{natbib}
\usepackage{url} 

\usepackage{amssymb, amsfonts, amsthm, bm}
\usepackage{mathtools, mathrsfs}
\usepackage{relsize} 
\usepackage{hyperref}

\newtheorem{definition}{Definition}

\newtheorem{lemma}{Lemma}

\newtheorem{example}{Example}

\def\beginmat{ \left( \begin{array} }
\def\endmat{ \end{array} \right) }

\DeclareMathOperator{\E}{E}

\DeclareMathOperator{\Cov}{Cov}

 \newcommand{\noop}[1]{}

\usepackage{color,soul}
\soulregister\cite7
\soulregister\citep7
\soulregister\eqref7
\soulregister\pageref7

\startlocaldefs
\numberwithin{equation}{section}
\theoremstyle{plain}

\endlocaldefs

\begin{document}

\begin{frontmatter}
\title{Nonparametric estimation of utility functions \thanksref{T1}}
\runtitle{Nonparametric estimation of utility functions}

\begin{aug}
\author{\fnms{Mengyang} \snm{Gu}\thanksref{addr1}\ead[label=e1]{mgu6@jhu.edu}}
\author{\fnms{Debarun} \snm{Bhattacharjya}\thanksref{addr2}\ead[label=e2]{debarunb@us.ibm.com}}
\author{\fnms{Dharmashankar} \snm{Subramanian}\thanksref{addr2}\ead[label=e3]{dharmash@us.ibm.com}}



\address[addr1]{Department of Applied Mathematics and Statistics, Johns Hopkins University, Baltimore, MD 
    \printead{e1} 
}

\address[addr2]{IBM T.J. Watson Research Center, Yorktown Heights, NY  
 \printead{e2,e3} 
 }
%

\end{aug}
\begin{abstract}
Inferring a decision maker's utility function typically involves an elicitation phase where the decision maker responds to a series of elicitation queries, followed by an estimation phase where the state-of-the-art is to either fit the response data to a parametric form (such as the exponential or power function) or perform linear interpolation. We introduce a Bayesian nonparametric method involving Gaussian stochastic processes for estimating a utility function. Advantages include the flexibility to fit a large class of functions, favorable theoretical properties, and a fully probabilistic view of the decision maker's preference properties including risk attitude. Using extensive simulation experiments as well as two real datasets from the literature, we demonstrate that the proposed approach yields estimates with lower mean squared errors. While our focus is primarily on single-attribute utility functions, one of the real datasets involves three attributes; the results indicate that nonparametric methods also seem promising for multi-attribute utility function estimation.
 \end{abstract}

 
%



\end{frontmatter}

\section{Introduction}

Making decisions under uncertainty using decision theory requires that beliefs about uncertainties be represented by probabilities and preferences over outcomes be summarized by utilities. The latter is often accomplished by assessing a decision maker's utility function over a single attribute (such as monetary units) or over multiple attributes, depending on the decision situation under consideration.

The vast literature on modeling utility functions describes schemes where the decision maker responds to elicitation queries; these responses are then subsequently used to infer the decision maker's utility function. We distinguish between these phases and refer to them as {\bf elicitation} and {\bf estimation} respectively. Posing elicitation questions in a specific form and then estimating the functional form that best represents the decision maker's preferences often go hand in hand.
In this paper, we contribute to the literature on estimation by introducing a Bayesian nonparametric approach for modeling utility functions. Specifically, we demonstrate the advantages of using a Gaussian stochastic process (henceforth GaSP) over the state-of-the-art, where parametric methods and the nonparametric method of linear interpolation are most popular. We show that linear interpolation corresponds to a subclass of our GaSP model using a specific covariance function in which the process turns to be a {\em Wiener process} (\cite{morters2010brownian}); moreover, parametric methods can be incorporated into the GaSP model through the mean function.

While the GaSP model has been popular in domains such as nonlinear regression and classification in machine learning (\cite{rasmussen:williams:2006}), spatial statistics (\cite{cressie1993}]) and computer model emulation and calibration (\cite{sacks1989}), it has not been applied to the problem of estimating utility functions, to the best of our knowledge.
There is some related work in the artificial intelligence literature on preference modeling with GaSP. For instance, \cite{chu2005preference} and \cite{guo2010gaussian} use GaSP for preference learning over a set of items. In contrast, our work is on estimating utility functions based on elicitation schemes from the operations research and management science literature (see e.g. \cite{farquhar1984}).

A fundamental challenge in utility elicitation is that the number of assessed queries is often very small, since elicitation burden increases significantly and responses become less reliable when more questions are posed.
The main contribution of this work is a robust estimation approach using GaSP models that reduces predictive errors when there are a small number of observations, in comparison with approaches such as parametric methods and linear interpolation. Furthermore, we provide a coherent model to accurately quantify the uncertainty associated with the utility elicitation process, as compared to the point predictions that are popular in the literature. Another benefit of such a coherent model is that we are able to assess risk attitudes of decision makers more accurately than current adopted methods.
Our proposed methods for utility elicitation are available online as an R package (\cite{gu2018robustgasp}).


The remainder of the paper is organized as follows. In sections 2 and 3, we introduce basic notation and terminology, placing our contribution in the context of related literature in operations research and management science on utility elicitation and estimation. Here we motivate our research effort and also discuss some theoretical advantages of nonparametric methods. In section 4, we introduce our GaSP formulation for utility function estimation and also show how to compute derivatives for assessing risk attitude. In section 5, we make a practical case for our proposed method by comparing it with prevalent methods for estimating single-attribute utility functions using a few assessed points on the utility function. We conduct several simulation experiments and also study a real-world dataset (\cite{abdellaoui2007loss}).
In section 6, we delve into estimating multi-attribute utility functions, demonstrating that GaSP performs better than the estimation methods that were deployed for a three-attribute dataset (\cite{fischer2000attribute2}). Finally, we conclude the paper in section 7.

\section{Uncertainty Quantification for Utility Elicitation}

Consider outcomes $\mathbf x$ in a separable metric space $\mathcal X$ (e.g. $\mathbb R^n$).
\cite{debreu1954representation} showed that preferences over  uncertain outcomes in such a space are complete, transitive and continuous in $\mathcal X$ iff there exists a continuous utility function representation $U: \mathcal X \to \mathbb R$.
Behavioral literature indicates that people tend to construct their preferences during the elicitation process, often providing inconsistent responses to queries (\cite{sarah2006construction}). The implication is that a decision maker's utility function $U$ should perhaps be considered an approximate representation of their preferences; for this and other reasons we shall discuss shortly, we make the following distinction:

\begin{definition}[Noise-free vs. noisy assessment]
When the decision maker answers all the preference elicitation questions consistently with the same underlying (and typically unknown) utility function, the assessment is said to be noise-free, otherwise it is noisy.
\end{definition}

The notion of a `true' underlying utility function can be viewed as a theoretical construct and one that has been discussed often in the literature. One way to model the uncertainty in preference elicitation responses is to include a random response error to a systematic component, either as additive error (Laskey and Fischer (1987)) or as random parameters of the utility function
(\cite{eliashberg1985measurement}, Fischer et al. (2000a)).
Another approach is to treat the utility function as inherently stochastic (Becker et al. 1963).
In practice, decision analysts are well aware of these issues and have devised measures to counter inconsistencies (see for example \cite{KeeneyRaiffa76}).

Making a noise-related distinction enables us to express different sources of uncertainty in the elicitation process. One source of uncertainty is prediction uncertainty, representing the system's uncertainty about the decision maker's utility at an unassessed $\mathbf x$. This uncertainty is present regardless of whether decision makers are consistent with their answers. The second source of uncertainty depends on the decision maker. When they consistently answer questions with the same underlying utility function, the elicited utility $u(\mathbf x)$ is identical to $ U(\mathbf x)$ at each assessed $\mathbf x$. We refer to an estimation method that agrees with the elicited utility at each assessed $\mathbf x$ as an {\bf interpolator}. It is ideal for a method to be an interpolator for a consistent decision maker. Note that the second source of uncertainty only appears when decision makers do not answer questions consistently, leading to noisy assessments.

%

Preference elicitation for decision making under risk is typically conducted using {\bf gambles}. Consider gamble $(\mathbf x_a, p; \mathbf x_b)$ that results in outcome $\mathbf x_a$ with probability $p$ and outcome $\mathbf x_b$ with probability $1-p$, where  $\mathbf x_a, \mathbf x_b \in \mathcal X$.
In an elicitation query, the decision maker must evaluate two or more gambles presented to them, e.g. they may be asked to compare gamble $(\mathbf x_a, p; \mathbf x_b)$ with the degenerate gamble $(\mathbf x_c)$. If the assessment is noise-free and if the evaluation is done by expected utility theory (EUT), they prefer the first gamble iff $V_{EUT}(\mathbf x_a, p; \mathbf x_b) = p U(\mathbf x_a)+(1-p)U(\mathbf x_b) \geq U(\mathbf x_c)$, for some underlying utility function $U$ (\cite{vNM1947}).
A subsequent estimation task must be performed to infer $U$ from the responses.


There is significant empirical evidence from the descriptive literature on prospect theory demonstrating that people tend to overweight low probabilities and underweight high probabilities, thus under prospect theory, the gamble $(\mathbf x_a,p;\mathbf x_b)$ evaluates to
$V_{PT}(\mathbf x_a, p; \mathbf x_b)= \omega (p) U(\mathbf x_a) +\omega (1-p)  U(\mathbf x_b)$, where $\omega (.)$ is a probability weighting function and the reference point is assumed to be $0$ (\cite{kahneman1979prospect,tversky1992advances}). Prospect theory also explains observed behavior such as loss aversion and diminishing sensitivity relative to the reference point.

\cite{bleichrodt2001} and others argue that descriptive violations of expected utility bias utility elicitations, therefore the design of elicitation schemes and estimation of utility functions should be conducted based on prospect theory rather than expected utility theory, resulting in more accurate estimates of $U$. We are sympathetic to this view but as we explain in the next subsection, our research goal is purely that of estimating $U$, provided elicitation responses in the following form:

\begin{definition}[Assessed tuples] Assessed tuples are assessments of points on the utility function, $\left(\mathbf x_i, u(\mathbf x_i) \right)$ for $i=1,\ldots,n$, $\mathbf x_i \in \mathcal X$.
\end{definition}

We propose a GaSP approach to estimating a utility function given assessed tuples as input.
Our method can thus be applied to different data sets and is agnostic to the underlying theory. In fact, we demonstrate GaSP estimation using elicitation schemes based on both expected utility theory and prospect theory. Furthermore, it is also agnostic to the generative structure of randomness in the case of probabilistic preferences, i.e. whether the data is generated by an additive noise model, or a utility function with random parameters, or indeed any stochastic utility model.
As mentioned earlier, we consider both noise-free and noisy assessments. Note that for noise-free assessments, assessed tuples $\left(\mathbf x_i, u(\mathbf x_i) \right) = \left(\mathbf x_i, U(\mathbf x_i) \right)$
for underlying utility function $U$.

We refer the reader to \cite{farquhar1984} for a comprehensive review of preference elicitation schemes involving single attributes; \cite{KeeneyRaiffa76} also describe elicitation schemes for multi-attribute problems.


\section{Utility Estimation}
\label{subset:est}
The goal of any estimation task associated with preference elicitation is to use responses to the elicitation queries to infer the decision maker's utility function $U(\mathbf x)$. Here we discuss parametric estimation and linear interpolation, highlighting potential limitations.

\subsection{Parametric Utility Estimation}

Parametric estimation is a popular approach for estimating utility functions involving a single attribute $x$ (\cite{eliashberg1985measurement, kirkwood2004approximating}).
The most common parametric forms are the exponential and power functions. The exponential utility function takes the form $a - b$ sgn$(\rho)$ exp $( -{x}/{\rho} ) $, where $a$ and $b>0$ are constants and sgn$(\rho)$ is the sign of the risk tolerance parameter $\rho \neq \infty$.
The power utility function is of the form $a + b$ sgn$(\alpha)$ sgn$(x)$ $|x|^{\alpha}$, with constants $a$ and $b>0$, where sgn$(\alpha)$ and sgn$(x)$ are the signs of $\alpha \neq 0$ and $x$.
Note that the power function exhibits discontinuous behavior around $0$ for $\alpha < 0$; see \cite{wakker2008explaining} for details.
The limiting cases for the exponential and power functions are the linear and logarithmic functions; these two families of functions are the only ones that satisfy constant risk aversion and constant relative risk aversion respectively (\cite{pratt1964}).

For multi-attribute utility functions, the most popular approach is to check for utility independence assumptions that decompose the multi-dimensional function into an additive or multiplicative function of one-dimensional marginal utility functions (\cite{KeeneyRaiffa76}). These marginal functions typically take the form of the afore-mentioned standard parametric models.
There has also been literature on utility functions that do not enforce utility independence assumptions.
To account for utility dependence, for instance, parametric forms have also been used for conditional utility functions, i.e. the utility of a subset of attributes given that the other attributes are set at fixed consequences (see for example \cite{kirkwood1976parametrically}).
A more recent approach is to use a copula (\cite{abbas2009multiattribute}), which has the benefit that the level of dependency between marginal utility functions can be represented in an easily interpreted parametric way.
In our GaSP model, utility dependence between attributes is modeled through a product covariance function, as discussed in section~\ref{subsec:model_GP}.

It is not hard to see that a parametric method will not be an interpolator unless the decision maker's underlying utility function follows the parametric class being used. The lack of this desirable property could result in large elicitation errors when the parametric class is misspecified, as we show in Section~\ref{sec:simulation}.

\subsection{Linear Interpolation}

An alternate approach that is popular in the empirical literature on estimating single-attribute utility functions is that of piece-wise linear interpolation across assessed tuples (\cite{abdellaoui2000parameter,abdellaoui2007loss}). Such an approach is essentially equivalent to the {\em predictive mean} of the extended Wiener process, defined as a stochastic process $W_t$ with independent, normally distributed increments $W_t-W_s \sim \mathcal N(0,t-s)$ for $t\geq s\geq 0$ with continuous sample paths (\cite{karlin1975}). A Wiener process is typically defined to have initial value $W_0=0$ but we relax this assumption. The predictive distribution formalized in the following lemma. (Please see Appendix for all proofs.)

 \begin{lemma}
Assume $W_t$, $t \in \mathcal T$ follows a Wiener process. Assume we have observations $W_{t_1},...,W_{t_n}$ with $0< t_1< t_2<...<t_n$. For any $t_{i} \leq t_*\leq t_{i+1}$, for any $1\leq i<n$ and $i \in \mathbb N$ , the predictive distribution of $W_{t_*}$ given  $W_{t_1},...,W_{t_n}$ is
 \[W_{t_*}| W_{t_1},...,W_{t_n} \sim \mathcal N(\mu_*, V_* ), \]
where $\mu_*=\frac{(t_{i+1}-t_*)W_{t_{i}}+ (t_*-t_i)W_{t_{i+1}} }{t_{i+1}-t_i}$ and $V_*=\frac{(t_{i+1}-t_*)(t_*-t_i)}{t_{i+1}-t_i}.$
%
\label{lemma:BM}
 \end{lemma}

Lemma~\ref{lemma:BM} states that if the utility function is modeled as a Wiener process for a single attribute $t$,  the posterior mean $\mu_*=\E[W_{t_*}| W_{t_1},...,W_{t_n}] $ for assessing the utility at a point $t_*$ (that has not been assessed) is equivalent to linear interpolation between two nearby assessed tuples.

 \begin{figure}[t]
\centering
  \begin{tabular}{cc}

\includegraphics[scale=.5]{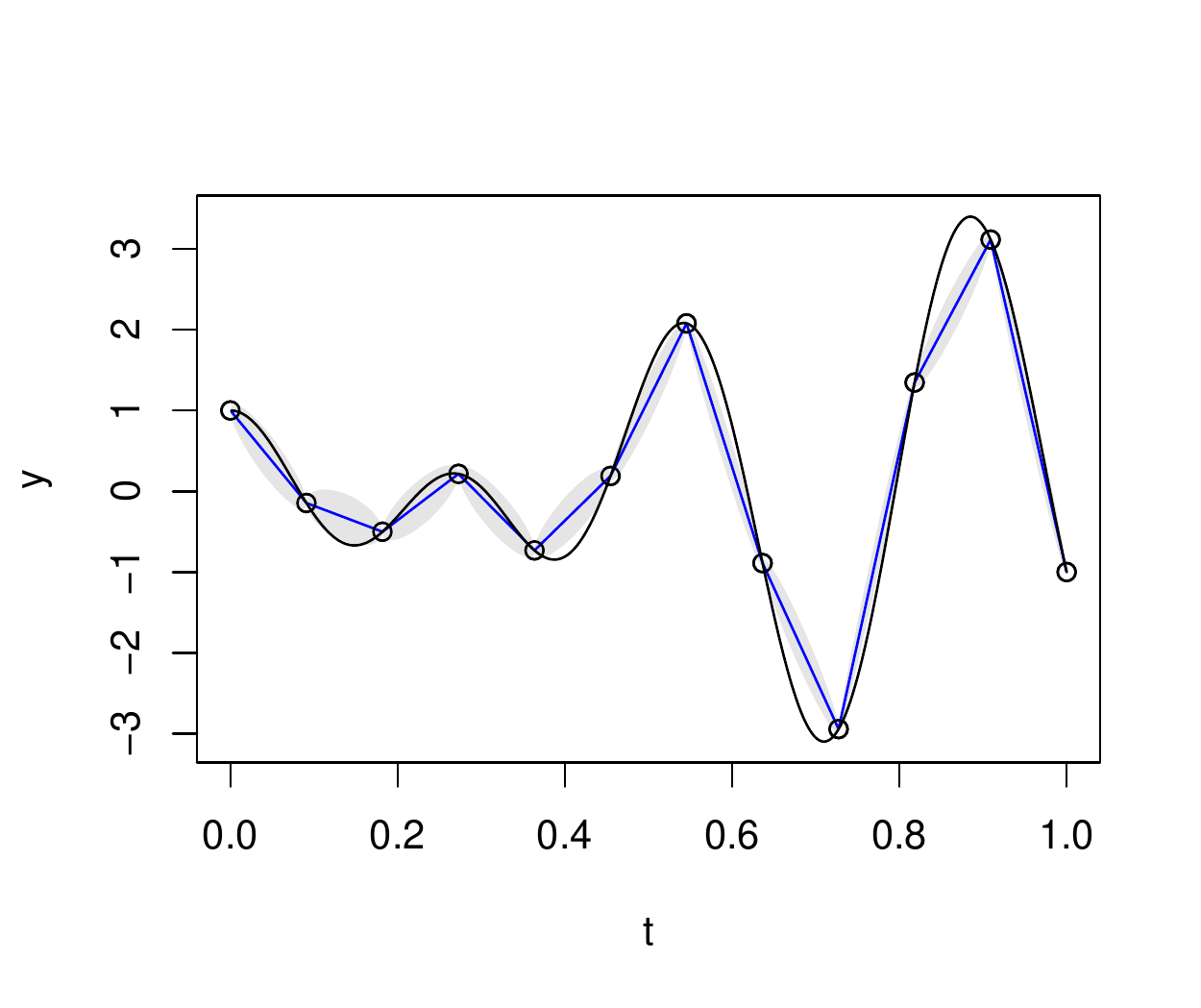}
\includegraphics[scale=.5]{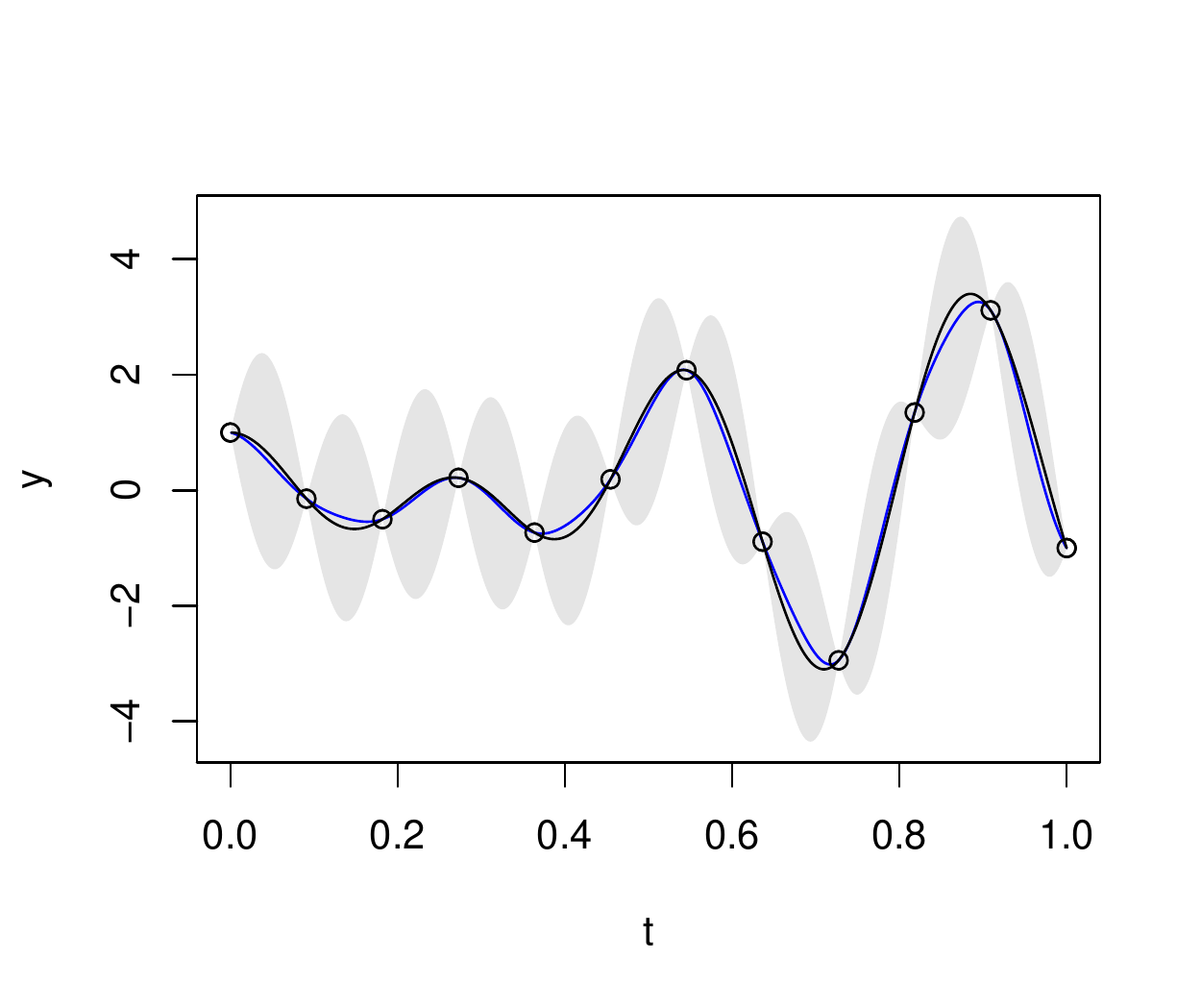}
  \end{tabular}

\caption{Interpolation of the function $y= 3sin(5\pi t)+cos(7\pi t)$ plotted as black curves with 12 assessments equally spaced in $[0,1]$ (black circles). Predictions by the extended Wiener process (left panel) and the GaSP model (right panel) with the default setting in the RobustGaSP R Package (\cite{gu2018robustgasp}) are plotted as blue curves . The 95\% predictive confidence intervals are shown by the grey area. }
\label{fig:interpolation}
\end{figure}

Indeed, a Wiener process is a special case of GaSP (which we will formally define in the next section) with initial value $W_0=0$, mean zero and covariance function $\Cov(W_s,W_t)=min(s,t)$ at any $s,t \geq 0$, and continuous sample path. However, a Wiener process is not differentiable everywhere and consequently using the posterior mean (i.e. linear interpolation) poses problems for estimating the coefficient of risk aversion.


\begin{example}[Linear interpolation and overconfidence]
Consider Figure~\ref{fig:interpolation}, which displays the function $y= 3sin(5\pi t)+cos(7\pi t)$ (treated as unknown) whose values are assessed at 12 equally spaced points in $[0,1]$.
The left panel shows the prediction (blue curves) by the extended Wiener process (for which we do not assume $W_0=0$). Not only does the prediction show discrepancy at places where the derivatives of the function change, it is also clearly overconfident as the 95\% confidence interval covers regions that are a lot smaller than the nominal 95\%. In comparison, the right panel is the prediction by the method we propose with the same 12 assessed points, using the default setting in the RobustGaSP R Package (\cite{gu2018robustgasp}). The improvement arises due to the use of other covariance functions with objective Bayesian inference, discussed in the next section in detail. While this example illustrates a shortcoming of linear interpolation using a generic function that is measured at a finite number of values in its domain, the above weakness is relevant for our purposes of estimating a utility function using experimentally assessed tuples.
\end{example}



Another limitation of a Wiener process estimation approach is that it is only defined in a one-dimensional domain and thus is limited to single-attribute utility function estimation. Although there is some literature on interpolation in multi-attribute problems (see for example \cite{bell1979multiattribute}), it is a challenging task due to the curse of dimensionality. A more general GaSP approach with suitable covariance functions built on the space of multiple attributes may be able to better handle the estimation task.

\subsection{Quantile-Parameterized Distributions}

Quantile-parameterized distributions (QPD) have recently been introduced for modeling uncertainties in decision analysis (\cite{keelin2011quantile,hadlock2017johnson}). This approach characterizes a continuous probability distribution based on a number of assessed quantile/probability pairs.
Although QPDs have not been discussed in the context of utility elicitation, it is straightforward to apply the approach conceptually as assessed tuples are analogous to quantile/probability pairs.
Denoting these assessed tuples as $(x_i, u_i)$ for $i=1,..,n$, where $u_i$ is the assessed utility scaled from $[0,1]$, the inverse CDF of a QPD takes the following form:
\[F^{-1}(u)=\left\{\begin{matrix}
\hspace{-.6in} L_0  & u=0,\\
\sum^q_{i=1}  a_ig_i(u) & \hspace{.25in} 0<u<1, \\
\hspace{-.6in} L_1 & u=1,
\end{matrix}\right.\]
with left handed limit $L_0=\lim\limits_{y\to 0^+}F^{-1}(u)$, right handed limit $L_1=\lim\limits_{y\to 1^-}F^{-1}(u)$, and $g_i(\cdot)$ referring to basis functions for $i=1,...,q$, $q\leq n$. When the domain of $x$ is a real line, a simple choice is the Q-normal distribution, where the basis functions are $g_1(u)=1$, $g_2(u)=\Phi^{-1}(u) $, $g_3(u)=y\Phi^{-1}(u)$, $g_4(u)=u$, with $\Phi$ being the normal CDF (\cite{keelin2011quantile}). Note that the results depend entirely on the choice of basis functions; further research is required to explore suitable models for utility elicitation.
Here we consider QPDs solely as another benchmark for the GaSP model since they provide some flexibility to model a curve ranging from $[0,1]$.

\section{Estimation with GaSP}

We introduce a Bayesian nonparametric method that takes assessed tuples as training data input, regardless of the underlying theory and assumptions used to derive them, and provide an estimated utility function $\hat{u} ( \mathbf x )$, where $\mathbf x$ could either be a single attribute or multiple attributes. Bayesian inference explicitly provides a formal way to quantify the uncertainty introduced by the estimation through the likelihood and prior, as discussed in detail in this section.

\subsection{Model Formulation}
\label{subsec:model_GP}

To set notation, let $\mathbf x=(x_1,...,x_p)^T$ be a vector of $p$ different attributes and let  $u(\mathbf x)$ be the utility evaluated at $\mathbf x$. Let us consider a random utility function modeled in a general regression way with the following form,
   \begin{equation}
    u(\mathbf x)= m(\mathbf x)+z(\mathbf x),
   \label{equ:randutility}
   \end{equation}
where $m(\mathbf x)$ is the {\bf mean function}, modeled as
    \[ E [u( \mathbf x)]= m({\mathbf x})=\mathbf h( \mathbf x)\bm \theta =\sum^q_{j=1}  h_j({\mathbf x})\theta_j,\]
where $\mathbf  h(\mathbf x)$ is assumed to be a $q$ dimensional domain dependent basis function for any $ \mathbf x \in \mathcal X$, with unknown regression parameters $\theta_j$ for each basis function $ h_j({\mathbf x})$. $h_j({\mathbf x})$ could be chosen, e.g., as a particular parametric form as specified in Section~\ref{subset:est} or as a polynomial function in $\mathbf x$. For the additive residual term, instead of taking $z(\mathbf x)$ as independent measurement errors as in \cite{eliashberg1985measurement}, we model $z(\cdot)$ as a stationary GaSP
\begin{equation}
 z(\cdot)  \sim  GaSP(0, \,\sigma^2c(\cdot,  \cdot) ) \,,
\label{equ:gp}
\end{equation}
with variance  $\sigma^2$ and the pair-wise correlation function $c(\cdot, \cdot)$. In return, the joint distribution of any $n$ inputs $\{ \mathbf{x}_1,\ldots,\mathbf{x}_n\} \in  \mathscr{X} $, follows a multivariate normal distribution,
\begin{equation}
 \big((z(\mathbf{x}_1),\ldots,z(\mathbf{x}_n))^T\mid (\sigma^2, \,{\mathbf C}) \big) \sim \mathcal{N} (\mathbf 0, \sigma^2   {\mathbf C}  )\,,
 \label{equ:multinormal}
 \end{equation}
i.e. a normal distribution that is conditional on the unknown variance $\sigma^2$ and the Gram matrix $\mathbf C$ (\cite{rasmussen:williams:2006}) whose $(i,j)$ element is $c \, (\mathbf x_i, \mathbf x_j)$.
By definition, the covariance of the utility is
  \[ \Cov(u( \mathbf x_a), u( \mathbf x_b))=\sigma^2c( \mathbf x_a, \mathbf x_b ),\]
for any $\mathbf x_a, \mathbf x_b \in \mathcal X$. If $p=1$ (i.e. single attribute) and $c(x_a,x_b)= min(x_a, x_b)$, GaSP becomes a Wiener Process when the process is defined on $x \in [0,+\infty)$, with initial value $0$. To extend the definition to the case when $p>1$, the  isotropic assumption is sometimes made for modeling a spatial process (\cite{gelfand2010handbook}), meaning that the correlation function $c(\mathbf x_a, \mathbf x_b )$ is only a function of $||\mathbf x_{a}- \mathbf x_{b}||$  where  $||\cdot||$ is the Euclidean distance. However, the domain of attributes typically varies on completely different scales (e.g. between the price and comfort of a car), so the effect of the attributes on the correlations will be highly variable. Consequently, the assumption of isotropy may not be reasonable. Instead, the product correlation function is often assumed:
\begin{equation}
c(\mathbf x_a, \, \mathbf x_b)=\prod^p_{l=1}c_l( x_{al},  x_{bl}),
\label{equ:product_c}
 \end{equation}
where $c_l(\cdot,\,\cdot)$ is a one-dimensional correlation function for the $l^{th}$ attribute.  We list several frequently used correlation functions in Table \ref{tab:corrfun}. The difference between the above product correlation function and the isotropic assumption is that for the former, there are parameter(s) in each correlation  $c_l(\cdot,\,\cdot)$ that can control the smoothness of the utility function on this attribute (which could be learned from the data), while  the latter assumes the Euclidean distance.


\begin{table}[t]
\begin{center}
\begin{tabular}{lr}
  \hline
                        & $c_l(d_l)$   \\
  \hline
  Power Exponential              & $ \exp\{-(d_l/\gamma_l)^{\nu_l} \}$, $\nu_l \in (0,2]$ \hfill \\
   Spherical                &$\left( 1-\frac{3}{2}\left( \frac{d_l}{\gamma_l} \right) + \frac{1}{2}\left( \frac{d_l}{\gamma_l} \right)^3 \right) \mathbf 1_{[d_l/\gamma_l\leq 1]}$    \hfill \\
Rational Quadratic       & $\left(1+ \left( \frac{d_l }{\gamma_l} \right)^2 \right)^{-\nu_l},\, \nu_l  \in (0,+\infty)$    \hfill \\
Mat\'{e}rn   &  $\frac{1}{2^{\nu_l-1}\Gamma(\nu_l)}\left(\frac{d_l}{\gamma_l} \right)^{\nu_l} \mathcal K_{\nu_l} \left(\frac{d_l}{\gamma_l} \right), \, \nu_l  \in (0,+\infty)$  \hfill \\
   \hline
\end{tabular}
\end{center}
   \caption{Popular choices of  correlation functions, when $c_l( x_{al},  x_{bl}) = c_l(d_l)$ with $d=|{x}_{al}-{x}_{bl}|$ for  $l=1,...,p$. Here $\nu_l$ is the roughness parameter, $\gamma_l$ is the range parameter,  $\Gamma(\cdot)$ is the gamma function and $\mathcal{K}_{\nu_l}(\cdot)$ is the modified Bessel function of second kind of order $\nu_l$. }
 \label{tab:corrfun}
\end{table}%

The power exponential covariance and the Mat\'{e}rn covariance have been used in many applications. When $\nu_l=(2k+1)/2$ where $k\in \mathbb N$,  Mat\'{e}rn correlation has a closed form. For example, when the roughness parameter  $\nu_l=5/2$, the Mat{\'e}rn correlation is

\begin{equation}
c_{Mat}(x_{al},x_{bl})=\left(1+\frac{\sqrt{5}d_l}{\gamma_l}+\frac{5d_l^2}{3\gamma_l^2}\right)exp\left(-\frac{\sqrt{5}d_l}{\gamma_l}\right),
\label{equ:matern2_5}
\end{equation}
where $d_l=|x_{al}- x_{bl}|$.  GaSP with the Mat{\'e}rn correlation defined in Equation (\ref{equ:matern2_5}) means it is  twice mean square differentiable (\cite{rasmussen:williams:2006}), allowing one to infer the risk attitude using derivatives of the GaSP as discussed later in Section~\ref{subsec:model_GP_derivative}. We use the Mat{\'e}rn correlation defined in Equation (\ref{equ:matern2_5}) for the purpose of demonstration but we do not preclude the use of other correlation functions with suitable differentiable results in future applications. In practice, the roughness parameter ($\nu_l$) is fixed at a chosen value (e.g. 5/2 for the Mat{\'e}rn), and the correlation function $c_l(.,.)$ is essentially parameterized by the range parameters ($\gamma_l$).

\subsection{Posterior Predictive Distribution}

Denote assessed tuples as $\left( \mathbf x^{\mathcal D}, \mathbf{u}(\mathbf x^{\mathcal D}) \right)$, where $\mathbf x^{\mathcal D} = \left\{\mathbf x^{\mathcal D}_1, \mathbf x^{\mathcal D}_2,\dots, \mathbf x^{\mathcal D}_n \right\}$ are $n$ points in the domain of the multi-attributes. As we are empirically limited by a small number of assessed utility values, we seek the {\bf posterior predictive distribution} of the utility function $u^*(x^*)$ at any input $ x^* \in \mathcal X$ based on the assessed tuples $\left( \mathbf x^{\mathcal D}, \mathbf{u}(\mathbf x^{\mathcal D}) \right)$. For simplicity, denote $\mathbf u^{\mathcal D}= \left(u(\mathbf x_1^{\mathcal D}),u(\mathbf x_2^{\mathcal D}),...,u(\mathbf x_n^{\mathcal D})\right)^T$ as the assessed utility points in the design. As per the chosen GaSP model, the likelihood is a multivariate normal likelihood:
  \[\mathcal L(\mathbf u^{\mathcal D}| \bm \theta,\sigma^2,\bm \gamma )=(2\pi\sigma^2)^{-n/2}|\mathbf C|^{-1/2} \exp\left\{- \frac{( \mathbf u^{\mathcal D}- \mathbf h(\mathbf x^{\mathcal D}) \bm \theta )^T (\sigma^2 \mathbf C)^{-1} ( \mathbf u^{\mathcal D}- \mathbf h(\mathbf x^{\mathcal D}) \bm \theta )}{2}  \right\},  \]
where $\mathbf{h}(\mathbf{x}^{\mathcal D} )$ is the $n \times q$  basis design matrix with $(i,j)$ element $h_j(\mathbf{x}^{\mathcal D}_i)$.  The model parameters for posterior estimation are the mean parameter $\bm \theta=(\theta_1,\theta_2,...,\theta_q)^T$, variance parameter $\sigma^2$ and range parameters $\bm \gamma=(\gamma_1,\gamma_2,...,\gamma_p)^T$ in the correlation function in Equation (\ref{equ:matern2_5}).

 We complete the model by specifying reference priors, $\pi$, for the model parameters (\cite{berger2001objective,paulo2005default,Gu2018robustness})
\begin{equation}
 {\pi }(\bm{\theta} ,{\sigma}^2,\bm {\gamma} ) \propto  \frac{|{\mathbf I^{*}}(\bm{\gamma} ){|^{1/2}}}{{\sigma }^{2}} \,,
 \label{equ:refprior}
 \end{equation}
where $\mathbf I^{*}(\cdot)$ is the expected Fisher information matrix, given as
\begin{equation}
{\mathbf I^*}({\bm \gamma} ) = {\left( {\begin{array}{*{20}{c}}
   {n - q} & {tr({\mathbf W_1})} & {tr({\mathbf W_2})} & {...} & {tr({\mathbf W_p})}  \\

   {} & {tr(\mathbf W_1^2)} & {tr({\mathbf W_1}{\mathbf W_2})} & {...} & {tr({\mathbf W_1}{\mathbf W_p})}  \\
   {} & {} & {tr(\mathbf W_2^2)} & {...} & {tr({\mathbf W_2}{\mathbf W_p})}  \\
   {} & {} & {} &  \ddots  &  \vdots   \\
   {} & {} & {} & {} & {tr(\mathbf W_p^2)}
 \end{array} } \right)_{(p+1) \times (p+1)}},
 \label{equ:efi}
 \end{equation}
where $\mathbf {W}_l = {\dot {{\mathbf C}}_l} \mathbf {Q}, \, 1\leq  l \leq p$ with $\mathbf {Q} = {\mathbf { C}^{ - 1}} ( \mathbf I_n - {{\mathbf h}(\mathbf{x}^{\mathcal D} )}{\{{{\mathbf h}^T(\mathbf{x}^{\mathcal D}  )} { {\mathbf { C} }^{ - 1}}{{\mathbf h(\mathbf{x}^{\mathcal D}  )}}\}^{ - 1}}{{{\mathbf h}^T(\mathbf{x}^{\mathcal D}  )}}{{\mathbf { C}}^{ - 1}})$ and $\dot {{\mathbf C}}_l$ is the derivative of $\mathbf C$ with regard to $\gamma_l$.


 After integrating $(\bm \theta, \sigma^2)$ using the above reference prior,  the marginal posterior of $\bm \gamma$ follows
\begin{equation}
 p({\bm \gamma }|\mathbf{u}^{\mathcal D})  \propto {\mathcal{L}}(\mathbf u^{\mathcal D}|\bm{\gamma })|{\mathbf I^{*}}(\bm \gamma ){|^{1/2}},
 \label{equ:iml}
 \end{equation}
where the marginal likelihood is
\begin{equation*}
\mathcal L(\mathbf u^{\mathcal D} |{\mathbf \gamma } ) \propto  |{\mathbf C}|^{- \frac{1}{2}} (|{{\mathbf h^T(\mathbf{x}^{\mathcal D}  ) }{ \mathbf C^{ - 1}}{\mathbf h(\mathbf{x}^{\mathcal D}   )} }|)^{- \frac{1}{2}}  \left( S^2\right)^{-(\frac{n - q}{2})},
\label{equ:mp}
\end{equation*}
 with $S^2=   { ( \mathbf{u}^{\mathcal D} )^T{\mathbf { Q} } {\mathbf{u}^{\mathcal D}  }  }$.

The predictive distribution from a Bayesian approach is obtained by integrating the uncertainty from the parameters
  \begin{equation}
  p(u(\mathbf x^*)| \mathbf u^{\mathcal D}) =\int \mathcal L(u(\mathbf x^*)|  \mathbf u^{\mathcal D}, \bm \theta,\sigma^2,\bm \gamma )p(\bm \theta,\sigma^2,\bm \gamma|  \mathbf u^{\mathcal D}  ) d\theta d\sigma^2 d\bm \gamma.
  \label{equ:pred_dist}
  \end{equation}

Note that $\bm \gamma$ cannot be marginalized out explicitly  - there is no prior on $\bm \gamma$ that leads to a closed form marginal likelihood after integrating out $\bm \gamma$. Numerical integration algorithms are both inefficient (as the computation requires the inverse of the correlation $\mathbf R$, which has $O(n^3)$ operations) and less stable. Instead, $\bm \gamma$ is often estimated by the  maximum marginal posterior mode
 \begin{equation}
 ({\hat \gamma}_1, \ldots {\hat \gamma}_p)=  \mathop{arg max }\limits_{ \gamma_1,\ldots, \gamma_p} \left\{ p({\bm \gamma }|\mathbf{u}^{\mathcal D}) \right\}.
 \label{equ:est_gamma}
 \end{equation}
As discussed in \cite{Gu2018robustness}, this leads to a robust estimation of $\bm \gamma$, yielding much better results than a non-robust method such as the maximum likelihood estimator (MLE). Note that the derivative of the reference prior is computationally intensive when the sample size is large. One can use some other objective priors that both have the robust estimation properties and the closed form derivatives (see e.g. \cite{gu2018jointly,Gu2016thesis}).


With the above setup,  the predictive distribution of $u(\mathbf x^*)$ at a new point $\mathbf x^* \in \mathcal X$, given the assessed tuples and the estimated range parameter $\hat{ \gamma}$, is a t-distribution (\cite{berger2001objective})
\begin{equation}
u(\mathbf x^*) \mid \mathbf{u}^{\mathcal D}  ,{\hat { \gamma}} \sim t( \hat u({ \mathbf x }^{*}),\hat{\sigma}^2c^{**},n - q).
\label{equ:predictiongp}
\end{equation}
with $n-q$ degrees of freedom, where   $\hat{u} ({ \mathbf x }^{*})$, $\hat{\sigma}^2$ and $c^{**}$ are  given respectively as,
\begin{align*}
\hat{u} ({ \mathbf x}^{*}) =& { \mathbf h({ \mathbf x}^{*})} \hat{\bm{\theta}}+\mathbf{c}^T({\mathbf x}^*){{ {\mathbf C}}}^{-1}\left(\mathbf{u}^{\mathcal D}  -{{\mathbf{h}(\mathbf{x}^{\mathcal D}  )}}\hat{\bm{\theta}}\right) \,,\\
\hat{\sigma}^2 =& (n-q)^{-1}{\left( \mathbf{u}^{\mathcal D}  -{\mathbf{h}(\mathbf{x}^{\mathcal D}  )}\hat{\bm{\theta}}\right)}^{T}{{ {\mathbf C}}}^{-1}\left(\mathbf{u}^{\mathcal D} -{{\mathbf{h}}(\mathbf{x}^{\mathcal D}  )}\hat{\bm{\theta}}\right), \\
 c^{**} =&  c({ \mathbf x^{*}}, {\mathbf  x^{*}})-{ \mathbf{c}^T({\mathbf x}^*){ {{  {\mathbf C}} }}^{-1}\mathbf{c}({\mathbf x}^*)} + \left({{\mathbf h( \mathbf x^{*})}-{\mathbf{h}^T(\mathbf{x}^{\mathcal D}  )}{ {\mathbf C}}^{-1}\mathbf{c}({ \mathbf x}^*) }\right) ^T \\
 & \times \left(\mathbf{h}^T(\mathbf{x}^{\mathcal D} ){{ {\mathbf C}}}^{-1}{\mathbf{h}(\mathbf{x}^{\mathcal D}  )}\right)^{-1}\left({{\mathbf h(\mathbf  x^{*})}}-{\mathbf{h}^T(\mathbf{x}^{\mathcal D}  )} {{\mathbf C}}^{-1}\mathbf{c}({\mathbf x}^*) \right) \,, \nonumber
\end{align*}
with $\hat{\bm{\theta}}=\left({{\mathbf{h}}^{T}(\mathbf{x}^{\mathcal D}  )}{{ {\mathbf C}}}^{-1} \ {{\mathbf{h}}(\mathbf{x}^{\mathcal D}  )}\right)^{-1}{\mathbf{h}^T(\mathbf{x}^{\mathcal D}  )}{{ {\mathbf C}}}^{-1}{\mathbf{u}^{\mathcal D} }$ as the generalized least squares estimator for $\bm \theta$ and    $\mathbf{c}({\mathbf x}^*) = (c({\mathbf x}^*,\mathbf{x}^{\mathcal D}_1 ), \ldots, c({\mathbf x}^*,\mathbf{x}^{\mathcal D}_n ))^T$ being the correlation of the GaSP at ${ \mathbf x}^{*}$ and $\mathbf u^{\mathcal D}$.

The predictive mean, $\hat u(\mathbf x^*)$ is typically used for estimation of the utility function. The following lemma specifies that such an estimation is an interpolator.

\begin{lemma}[GaSP Interpolator]
\label{lemma:GaSP_interpolator}
The predictive mean $\hat u(\mathbf x^{\mathcal D}_i) = U(\mathbf x^{\mathcal D}_i)$ for any $i=1,...,n$ for noise-free assessments with utility function $U(\cdot)$.
\end{lemma}

For noisy assessments, we do not expect the prediction of GaSP to be exact at the assessed points.
To model such situations, an independent noise can be added in Equation (\ref{equ:randutility}) by defining $\tilde z(\mathbf x)= z(\mathbf x) +\varepsilon$, where $\varepsilon$ is independent white noise. Objective Bayesian inference for such a GaSP model is similar to the method discussed above (\cite{ren2012objective,Gu2016PPGaSP}). The afore-mentioned methods have been developed as an R package (\cite{gu2018robustgasp}).


\subsection{Derivatives of GaSP}
\label{subsec:model_GP_derivative}
Differentiability is an important property of utility functions, e.g., the Arrow-Pratt measure of local risk aversion is given as $\lambda(x) = -u''(x)/u'(x)$. Our proposal to use GaSP is helpful in this regard since the derivative processes are also GaSP when the covariance function is mean square differentiable (\cite{rasmussen:williams:2006}).

Closed form expressions of derivatives of GaSP with Gaussian correlation (i.e. power exponential correlation with $\nu_l=2$ in Table \ref{tab:corrfun}) are shown in \cite{wang2012bayesian} (chapter 3). Here we present the derivatives of the Mat{\'e}rn class correlation with roughness parameter equal to 2.5. For simplicity, we assume there is only a single attribute in this subsection, but the following result can be easily extended to directional derivative processes with regard to each attribute, i.e. $\frac{\partial u(\mathbf x)}{\partial x_l}$.

Denote $\mathbf d^*=(|x^{\mathcal D}_1-x^*|,\ldots, |x^{\mathcal D}_n-x^*|)^T$ and $\mathbf g=(g_1,\ldots, g_n)^T$ with $g_i=\left\{\begin{matrix}
1 \quad x^{*}\geq x^{\mathcal D}_i\\
-1 \quad x^*<x^{\mathcal D}_i
\end{matrix}\right.$. Let $\circ$ denote the element-wise product of each entry between two matrices. The following two lemmas present the joint distribution of the GaSP and its derivatives, and they can be derived through directly differentiating the covariance function (see e.g. \cite{golchi2015monotone}).


\begin{lemma}
\label{lemma:1dev}
The joint distribution of GaSP ${\mathbf {u}^{\mathcal D}}$ and GaSP derivative ${ u'} ({ x}^{*})$ on any $x^* \in \mathcal X$ is:
	\begin{equation*}
\left( \begin{gathered}
  {\mathbf {u}^{\mathcal D}}  \hfill \\
  { u'} ({ x}^{*}) \hfill \\
\end{gathered}  \right) \mid \bm\theta ,{\sigma ^2},{\bm\gamma},   \sim N\left(\left( \begin{array}{*{20}{c}}
\mathbf  h({ {\mathbf x^{\mathcal D}}}){\bm\theta}   \\
    { \mathbf h'}( x^*  ){\bm\theta}  \\
 \end{array}\right) ,{\sigma ^2}\left( {\begin{array}{*{20}{c}}
   {  {\mathbf C} }    &\mathbf  c^{01}({    \mathbf x^{\mathcal D}, x^{*}})   \\
\mathbf  c^{10}(  x^{*},\mathbf x^{\mathcal D}) &  c^{11}({ x^{*}} , { x^{*}}) \\
 \end{array} } \right)\right).
 \end{equation*}
 where $\mathbf h'({ {x}}^{*})=\frac{\partial \mathbf h( x^{*})}{\partial x^{*} }$, $\mathbf c^{01}({  \mathbf  x^{\mathcal D}, x^{*}})=-\frac{5 \mathbf g }{3\gamma^2}\circ(\mathbf d^*+\frac{\sqrt{5} \mathbf d^* \circ \mathbf d^* }{\gamma})\circ exp(-\frac{\sqrt{5} \mathbf d^*}{\gamma}) $,   and $c^{11}({ x^{*}} , { x^{*}})= \frac{5}{3 \gamma^2}$, and $\mathbf c^{10}(  x^{*},\mathbf x^{\mathcal D}) =\mathbf c^{01}({    \mathbf x^{\mathcal D}, x^{*}})^T$.
\end{lemma}


\begin{lemma}
\label{lemma:2dev}
 The joint distribution of GaSP ${\mathbf {u}^{\mathcal D}}$ and GaSP derivative ${ u''} ({ x}^{*})$ on any $x \in \mathcal X$ is:
	\begin{equation*}
\left( \begin{gathered}
  {\mathbf {u}^{\mathcal D}}  \hfill \\
  { u''} ({ x}^{*}) \hfill \\
\end{gathered}  \right) \mid \bm\theta ,{\sigma ^2},{\bm\gamma},   \sim N\left(\left( \begin{array}{*{20}{c}}
   {\mathbf h}( \mathbf x^{\mathcal D}  ){\bm\theta}   \\
  \mathbf h''({ {x}}^{*}){\bm\theta}   \\
 \end{array}\right) ,{\sigma ^2}\left( {\begin{array}{*{20}{c}}
   {  {\mathbf C} }    & \mathbf c^{02}({   \mathbf x^{\mathcal D}, x^{*}})   \\
\mathbf  c^{20}( x^{*},\mathbf x^{\mathcal D}) &  c^{22}({ x^{*}} , { x^{*}}) \\
 \end{array} } \right)\right).
 \end{equation*}
 where $h''( x^{*})=\frac{\partial^2 h( x^{*})}{\partial x^{*}\partial x^{*} }$, $\mathbf c^{02}({   \mathbf x^{\mathcal D}, x^{*}})=\frac{5}{3\gamma^2}\left\{\frac{5\mathbf d^* \circ \mathbf d^*}{\gamma^2}-1-\frac{\sqrt{5}\mathbf d^*}{\gamma}\right\}\circ exp(-\frac{\sqrt{5} \mathbf d^*}{\gamma}), $ and $c^{22}( x^* ,  x^*)=\frac{25}{\gamma^4}$ and $\mathbf c^{20}( x^{*},\mathbf x^{\mathcal D}) = \mathbf c^{02}({   \mathbf x^{\mathcal D}, x^{*}})^T$.
\end{lemma}

Note that the above joint distribution is used to calculate the predictive distribution of the derivative. Indeed, by marginalizing out the mean and variance parameter, the predictive distribution of $u''({x}^{*})$ given $\mathbf{u} $ and  $\hat{\bm \gamma}=(\hat \gamma_1, \ldots, \hat \gamma_p)$, is a t-distribution

\begin{equation}
u''({ x^{*}}) \mid \mathbf{u}^{\mathcal D} ,{\hat {\bm \gamma}} \sim t( \hat u''({ x}^{*}),\hat{\sigma}^2c{''},n - q),
\label{equ:predictiongp_second_dev}
\end{equation}
with $n-q$ degrees of freedom, where
\begin{eqnarray*}
\hat{u}'' ({ x}^{*}) &=& { \mathbf h''({ x}^{*})} \hat{\bm{\theta}}+\mathbf{c}^{02}(\mathbf{x}^{\mathcal D}, {x}^*){{ {\mathbf C}}}^{-1}\left(\mathbf{u}^{\mathcal D} -{{\mathbf{h}(\mathbf{x}^{\mathcal D}  )}}\hat{\bm{\theta}}\right) \,,\\
\hat{\sigma}^2 &=&(n-q)^{-1}{\left(\mathbf{u}^{\mathcal D} -{\mathbf{h}(\mathbf{x}^{\mathcal D}  )}\hat{\bm{\theta}}\right)}^{T}{{ {\mathbf C}}}^{-1}\left(\mathbf{u}^{\mathcal D} -{{\mathbf{h}}(\mathbf{x}^{\mathcal D}  )}\hat{\bm{\theta}}\right), \\
 c{''} &=&  c^{22}({ x^{*}}, { x^{*}})-{\mathbf c^{02}({  \mathbf  x^{\mathcal D}, x^{*}}) { {{  {\mathbf C}} }}^{-1} \mathbf c^{20}({  \mathbf  x^{\mathcal D}, x^{*}}) } + \left({{\mathbf h''( x^{*})}-{\mathbf{h}^T(\mathbf{x}^{\mathcal D}  )}{ {\mathbf C}}^{-1}  \mathbf c^{20}({  \mathbf  x^{\mathcal D}, x^{*}})  }\right) ^T \\
&\hspace{+.3in}& \times \left(\mathbf{h}^T(\mathbf{x}^{\mathcal D} ){{ {\mathbf C}}}^{-1}{\mathbf{h}(\mathbf{x}^{\mathcal D}  )}\right)^{-1}\left({{\mathbf h''( x^{*})}}-{\mathbf{h}^T(\mathbf{x}^{\mathcal D}  )} {{\mathbf C}}^{-1}\mathbf c^{20}({  \mathbf  x^{\mathcal D}, x^{*}})  \right) \,. \nonumber
\end{eqnarray*}

The predictive distribution of $u'({x}^{*})$ is similar to Equation (\ref{equ:predictiongp_second_dev}). The point here is to demonstrate that the risk attitude can be obtained using the derivative processes with full assessment of the uncertainty associated with the analysis.

\section{Single Attribute Utility Function Estimation}

So far we have primarily seen the theoretical benefits of estimating utility functions with GaSP. In this section, we explore practical ramifications on estimating single-attribute utility functions by first performing simulation experiments and then analyzing a real dataset.

\subsection{Simulation Experiments}
\label{sec:simulation}

\subsubsection{Comparing utility function estimates}

In preference elicitation, only a limited number of questions can be posed, thus the number of assessed tuples $n$ is typically small:
$n=4$, $7$ and $10$ are considered for these experiments. Out of sample mean squared error (MSE),
$MSE={\sum_{i=1}^{n^*}{\left\{\hat u(x^*_i)-u(x^*_i)  \right\}^2}}/{n^*}$,
is utilized for comparison, where $x^*_i \in \mathcal X$ is the $i^{th}$ equally spaced held-out point and $n^*=1,001$ is used for testing throughout this section. We assume $U(x_{min})=0$ and $U(x_{max})=1$, where $x_{min}=0$ and $x_{max}=10^5$ are lower and upper bounds of the domain $\mathcal X$ in this subsection. We compare the exponential function (Exp), power function (Pow), linear interpolation (LI) and quantile-parameterized distribution (QPD) method with the GaSP model.

For the parametric methods and QPD method, we estimated the parameters with the minimum least squares error. For the QPD, we choose the basis function to be $g_1(u)=1$, $g_2(u)=\Phi^{-1}_*(u) $, $g_3(u)=u\Phi^{-1}_*(u)$, $g_4(u)=u$. As the domain of $x$ is $[0,10^5]$, the usual Q-normal distribution is not a sensible choice. Instead, we let $\Phi^{-1}_*(u)$ be a normal distribution truncated at $0$ and $10^5$ centered at $0$ with standard deviation $5\times 10^4$. This seems to perform the best among all basis functions considered.


In the first scenario, we assume that assessments are noise-free. Tables~\ref{tab:pow} and~\ref{tab:exp} display the out of sample MSE when the underlying utility function $U$ is a power function and an exponential function respectively. Because the  underlying utility function is assumed to be the power utility function in Table~\ref{tab:pow} and the power utility function is a subclass of models contained within the GaSP framework with mean basis $h(x)=x^{\alpha}$ and variance $\sigma=0$, we choose the mean function of the GaSP to be misspecified by selecting an inconsistent mean basis $(x^{0.5})$, demonstrating that GaSP performs reasonably well even in this scenario.


\begin{table}[ht]
\begin{center}
\small
\begin{tabular}{p{1cm}p{1.5cm}p{1.5cm}p{1.5cm}p{1.5cm}p{1.5cm}l}
  \hline
        Method          &$\alpha=0.7$&$\alpha=0.8$&$\alpha=0.9$ &$\alpha=1.5$&$\alpha=2.0$&$\alpha=2.5$  \hfill \\
  \hline

  Exp &$5.6\times 10^{-4}$&$2.0\times 10^{-4}$&$3.9\times 10 ^{-5}$&$3.3\times 10^{-4}$ &$6.2\times 10^{-4}$& $7.2\times 10^{-4}$   \hfill \\
  LI &  $4.6\times 10^{-5}$ & $2.3\times 10^{-5}$ &$6.1\times 10^{-6}$&$1.7\times 10^{-4}$ &$6.1\times 10^{-4}$ & $1.2\times 10^{-3}$   \hfill \\
  GaSP  &  {  $\bf 1.7 \times 10^{-6}$} &{  $ \bf 1.5 \times 10^{-6}$} &{  $\bf 5.3\times 10^{-7}$} &{  $\bf 2.2\times 10^{-5}$}&{  $\bf 4.9\times 10^{-5}$} &{  $\bf 1.7\times 10^{-6}$}  \hfill
  \\
  QPD   &  { $3.2 \times 10^{-5}$} &{  $2.5 \times 10^{-5}$} &{ $8.6\times 10^{-6}$} &{  $4.3\times 10^{-4}$}&{  $1.9\times 10^{-3}$} &{  $4.1\times 10^{-3}$}  \hfill \\

   \hline

  Exp &$5.6\times 10^{-4}$&$2.0\times 10^{-4}$&$3.9\times 10^{-5}$&$3.2\times 10^{-4}$&$6.0\times 10^{-4}$&$6.9\times 10^{-4}$   \hfill \\
  LI&  $9.5\times 10^{-6}$ & $5.0 \times 10^{-6}$ &$1.4 \times 10^{-6}$&$4.8\times 10^{-5}$&$1.9\times 10^{-4}$&$3.8\times 10^{-4}$    \hfill \\
  GaSP &  { $\bf 2.8\times 10^{-7}$} &{ $\bf 2.8\times 10^{-7}$} &{ $\bf 1.1\times 10^{-7}$} &{ $\bf 5.5\times 10^{-6}$}&{ $\bf 2.2\times 10^{-6}$}&{  $ \bf 5.0\times 10^{-6}$}  \hfill \\
  QPD &  { $2.2\times 10^{-5}$} &{ $ 1.7\times 10^{-6}$} &{ $5.7\times 10^{-6}$} &{ $ 2.8\times 10^{-4}$}&{ $1.2\times 10^{-4}$}&{  $ 2.8\times 10^{-3}$}  \hfill \\
\hline
\end{tabular}
\end{center}
   \caption{Out of sample MSE when $U$ is assumed to be power with different parameters $\alpha$. The number of observation is assumed to be $n=4$ for the first four rows and $n=7$ for the latter four rows.   }
   \label{tab:pow}
\end{table}

\begin{table}[ht]
\small
\begin{center}
\begin{tabular}{p{1cm}p{1.5cm}p{1.5cm}p{1.5cm}p{1.5cm}p{1.5cm}l}
  \hline
        Method           &$\rho=2.0$&$\rho=1.5$&$\rho=1.0$ &$\rho=-1.0$&$\rho=-1.5$&$\rho=-2.0$  \hfill \\
  \hline

  Pow   &$2.1\times 10^{-3}$&$1.1\times 10^{-3}$&$4.6\times 10^{-4}$&$2.5\times 10 ^{-4}$&$4.4\times 10^{-4}$ &$5.9\times 10^{-4}$   \hfill \\
  LI      & $ 1.1\times 10^{-4}$&  $4.8\times 10^{-5}$ & $1.7\times 10^{-5}$ &$1.7\times 10^{-5}$&$4.8\times 10^{-5}$ &$\bf 1.1\times 10^{-4}$    \hfill \\
  GaSP& { $4.7\times 10^{-4}$}  &  { $\bf 3.0 \times 10^{-6}$} &{  $\bf 9.6 \times 10^{-7}$} &{ $\bf 4.1\times 10^{-7}$} &{  $\bf 9.3\times 10^{-6}$}&{  $1.3\times 10^{-4}$}  \hfill \\
   QPD   & { $\bf 1.3\times 10^{-5}$}  &  { $1.4 \times 10^{-5}$} &{  $ 6.1 \times 10^{-5}$} &{ $3.7\times 10^{-5}$} &{  $2.0\times 10^{-4}$}&{  $6.6\times 10^{-4}$}  \hfill \\

   \hline

  Pow &$2.1\times 10^{-3}$&$1.1\times 10^{-3}$&$4.6\times 10^{-4}$&$2.5\times 10^{-4}$&$4.3\times 10^{-4}$&$5.9\times 10^{-4}$   \hfill \\
  LI& $1.9\times 10^{-5}$&  $7.6\times 10^{-6}$ & $2.6 \times 10^{-6}$ &$2.6 \times 10^{-6}$&$7.6\times 10^{-6}$&$1.9\times 10^{-5}$    \hfill \\
  GaSP& { $\bf 3.6\times 10^{-8}$ } &  { $\bf 1.9\times 10^{-8}$} &{ $ \bf 7.9\times 10^{-9}$} &{ $\bf 1.7\times 10^{-8}$} &{ $\bf 2.1\times 10^{-7}$}&{ $\bf 1.9\times 10^{-6}$}    \vspace{-.15in} \hfill \\
  QPD & { $1.2\times 10^{-5}$ } &  { $1.1\times 10^{-5}$} &{ $ 4.9\times 10^{-6}$} &{$2.5\times 10^{-5}$} &{ $ 1.3\times 10^{-4}$}&{ $4.3\times 10^{-4}$}  \vspace{-.15in}  \hfill \\

\hline
\end{tabular}
\end{center}
   \caption{Out of sample MSE when $U$ is assumed to be exponential with different parameters $\rho$. The number of observation is assumed to be $n=4$ for the first four rows and $n=7$ for the latter four rows. }
   \label{tab:exp}
\end{table}



Tables~\ref{tab:pow} and~\ref{tab:exp} both indicate that the GaSP method outperforms the other methods for most of the cases. The discrepancy indicated by MSE of testing points by GaSP is usually several orders of magnitude less than that for parametric fitting, and it is usually ten to hundred times better than LI. Note that we only use $n=4$ and $7$ (excluding two boundary points $x_{min}$ and $x_{max}$), which is reasonably small for elicitation purposes, yet the predictive power achieved by GaSP is already realized. The GaSP method with $4$ assessed points even performs better than its competitors with $7$ assessed points, meaning that the cost for eliciting the utility function can drop significantly using GaSP method.   Moreover, when the sample size increases, the MSE of GaSP decreases, while the MSE for parametric fitting does not necessarily decrease as it is not consistent when the class of parametric utility is misspecified.

The power and exponential families can sometimes be too restrictive, while GaSP is better for these methods because it allows a more flexible structure. The LI and QPD methods also induce a flexible structure. Indeed, they outperform GaSP for two cases when the true utility function is exponential and when the sample size is small ($n=4$), as the GaSP method uses a misspecified mean function. However, as the sample size increases to $n=7$ for these two cases, GaSP outperforms the other methods by a lot, meaning that GaSP model converges faster than the LI and QPD methods. A more reasonable mean function would, of course, increase the prediction of the GaSP model. However, as for all cases shown in Tables~\ref{tab:pow} and~\ref{tab:exp}, the mean function of the GaSP model is misspecified, but the GaSP model still shows the best prediction results for most cases. This feature makes it suitable for a default method for estimation of utility functions.

\begin{figure}[t]
\centering
  \begin{tabular}{cc}
	\includegraphics[width=.5\textwidth,height=.3\textwidth]{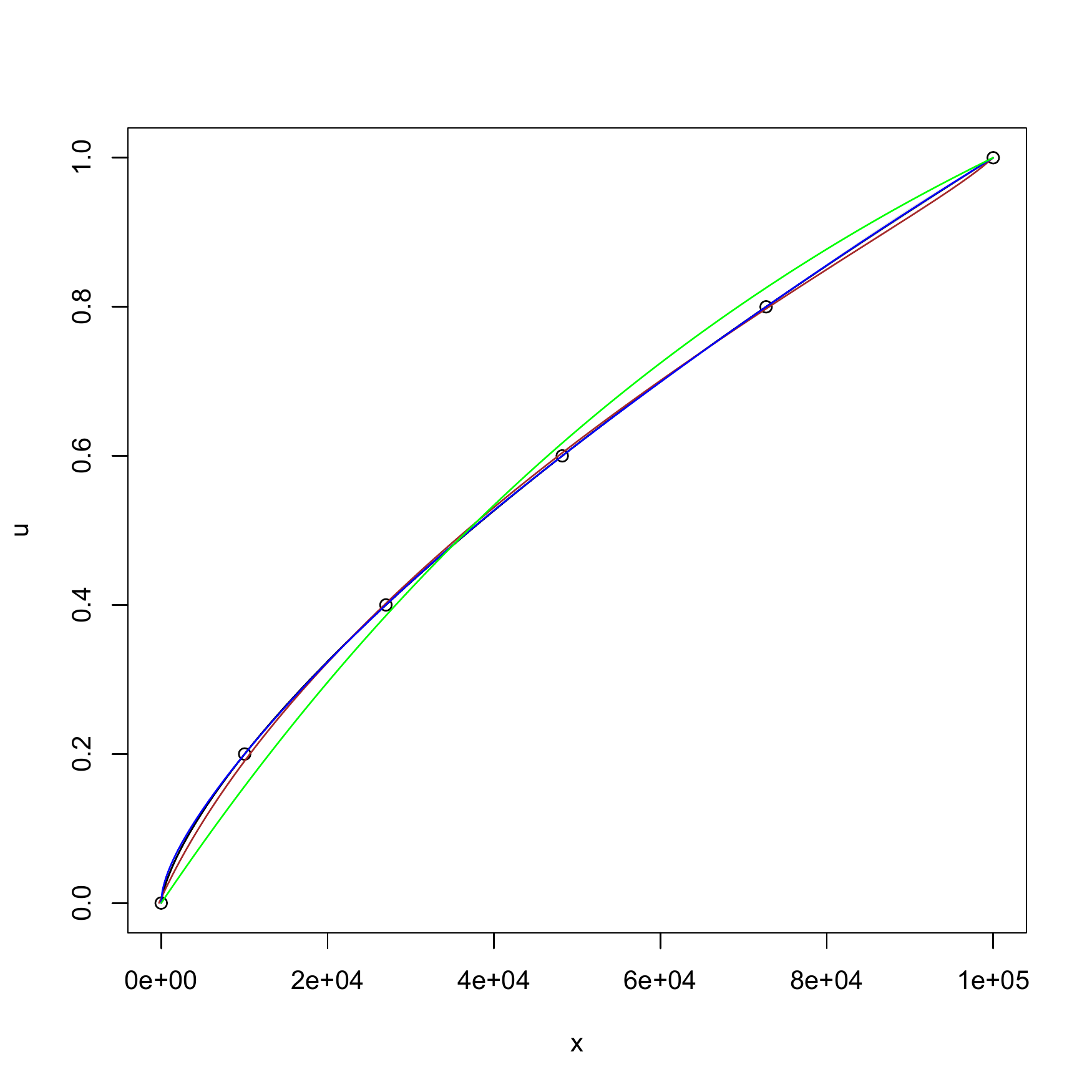}
	\includegraphics[width=.5\textwidth,height=.3\textwidth]{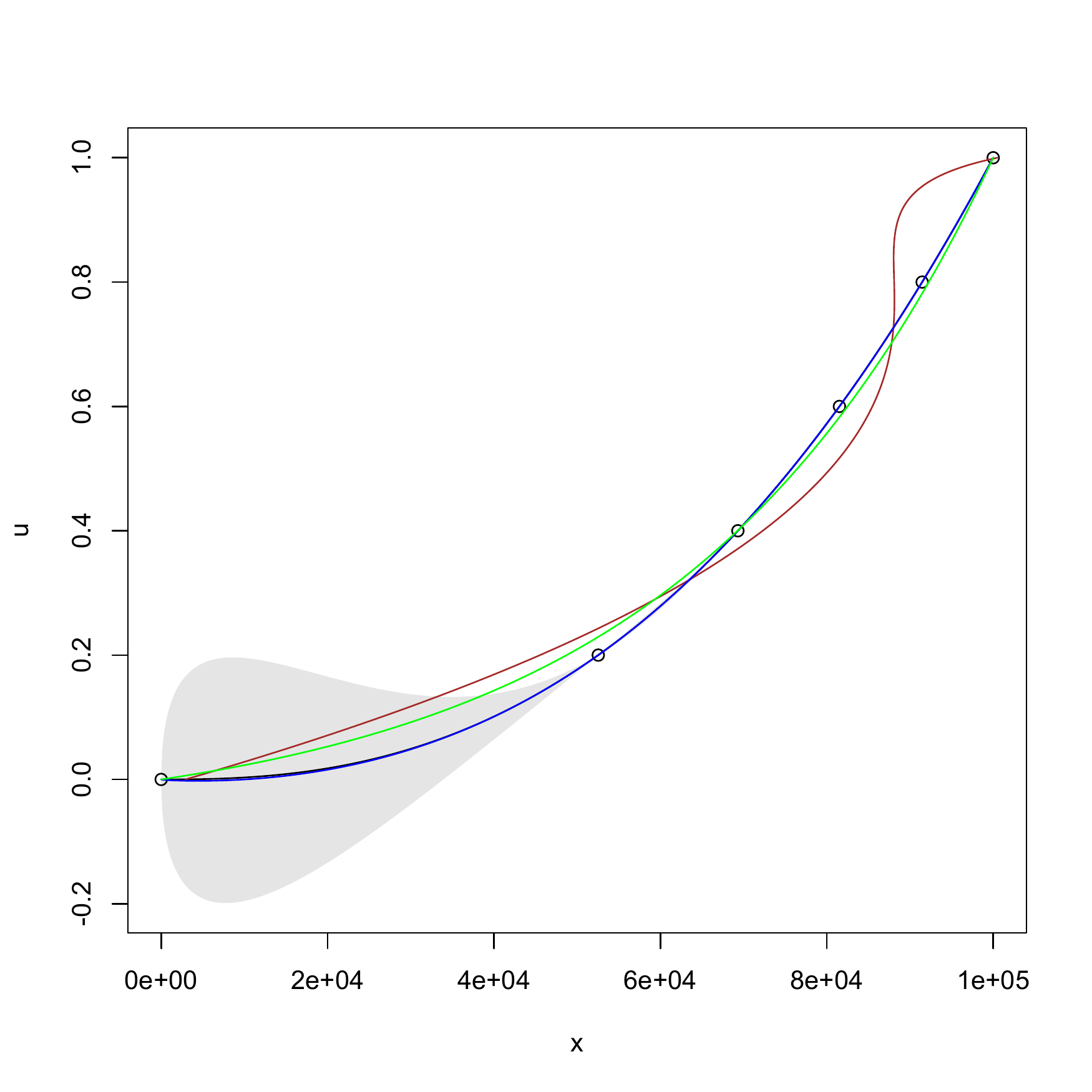} \\
	\includegraphics[width=.5\textwidth,height=.3\textwidth]{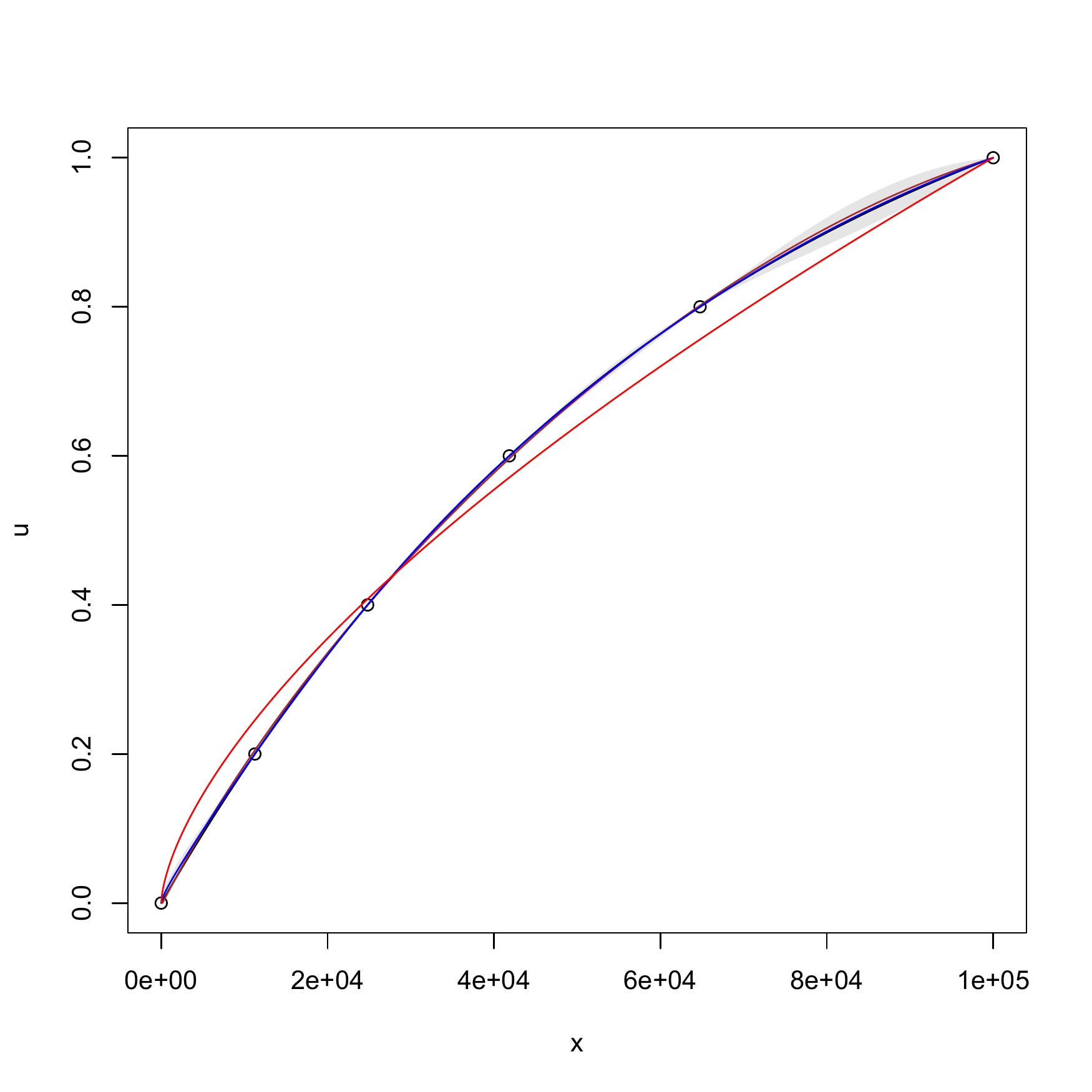}
	\includegraphics[width=.5\textwidth,height=.3\textwidth]{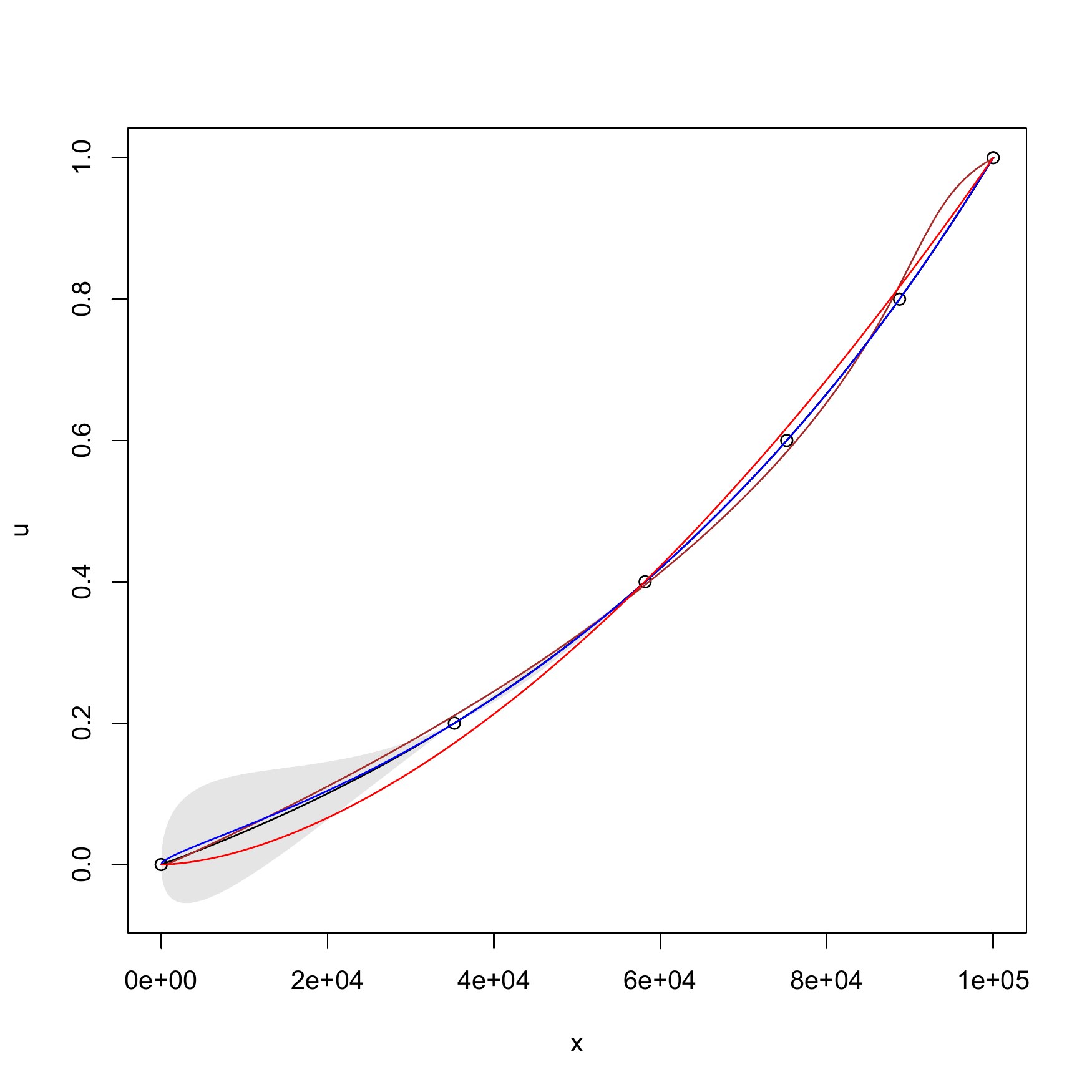}
  \end{tabular}
   \caption{Utility function estimation when the true utility function is power with $\alpha=.7$ (upper left panel),  $\alpha=2.5$  (upper right panel) and exponential with $\rho=1.5$  (lower right panel), $\rho=-1.5$ (lower right panel). Black circles represent assessed tuples and black curves are the true utility function.  Blue, brown, green and red curves are the fits by the GaSP model, QPD method, exponential and power functions, respectively. The shaded area are the 95\% predictive confidence interval by the GaSP model. For almost all cases, the blue and black curves overlap.}
 \label{fig:simulation}
\end{figure}

The fitted curves for the different methods and for 4 cases are plotted in Figure \ref{fig:simulation}. In almost all cases, the GaSP-fitted functions (blue curves) are close to the true functions (black curves). The GaSP model has very tight 95\% confidence intervals in most regions, meaning that it is confident about the prediction. The confidence interval of the GaSP is large when $x\in [0,4\times 10^4]$ and $x\in [0,2\times 10^4]$ in right panels in Figure~\ref{fig:simulation}, implying that the uncertainty of the GaSP model is high, and the prediction by the GaSP model has comparatively large discrepancy to the real utility function in these two regions. It is an advantage of the GaSP model that it has an internal assessment of the uncertainty, therefore one can determine whether and where to obtain more assessments given a required level of precision.

In the second scenario, we assume noisy assessments, specifically that utilities are random with additive noise,
 \begin{align}
  u(x)&=U(x) +\epsilon \nonumber\\
 \epsilon|\sigma^2_{\epsilon} &\sim \mathcal N(0,\sigma_{\epsilon}^2)
 \label{align:additive_noise}
 \end{align}
where $U(x)$ is the underlying utility function and $\sigma_{\epsilon}^2$ is the variance of the Gaussian noise. Since the assessed points are noisy, we simulated $N=200$ experiements to average out the design effect, and calculate the average MSE: $AvgMSE=\frac{1}{N}\sum_{j=1}^{N}MSE_j$. Here the total held-out testing points for each case is thus thus $n^*\times N=200,200$.

\begin{table}[ht]
\small
\begin{center}
\begin{tabular}{lp{1.5cm}p{1.5cm}p{1.5cm}p{1.5cm}p{1.5cm}l}
\hline
        Method          &$\alpha=0.7$&$\alpha=0.8$&$\alpha=0.9$ &$\alpha=1.5$&$\alpha=2.0$&$\alpha=2.5$  \hfill \\
  \hline

  Exp &$5.6\times 10^{-4}$&$2.0\times 10^{-4}$& $4.2\times 10^{-5}$ &$3.2\times 10^{-3} $&$6.0\times 10^{-4}$&$6.9\times 10^{-4}$ \hfill \\
  LI& $ 1.8 \times 10^{-5} $ & $1.7\times 10^{-5}$  & $1.5\times 10^{-5}$ &$3.4\times 10^{-5}$&$9.7\times 10^{-5}$&$1.9\times 10^{-4}$  \hfill \\
  GaSP&  $ {\bf 9.7\times 10^{-6}}$ &  ${\bf 9.1\times 10^{-6}}$   &${\bf 7.9\times 10^{-6}}$&${\bf 1.7\times 10^{-5}}$&${\bf 4.0\times 10^{-5}}$&${\bf 1.3\times 10^{-4}}$ \hfill \\
  QPD&  ${ 2.4\times 10^{-5}}$ &  ${ 1.9\times 10^{-5}}$   &${ 9.9 \times 10^{-6}}$&${ 2.1\times 10^{-4}}$&${ 9.2\times 10^{-4}}$&${ 2.1\times 10^{-3}}$ \hfill \\
  \hline
        Method           &$\rho=2.0$&$\rho=1.5$&$\rho=1.0$ &$\rho=-1.0$&$\rho=-1.5$&$\rho=-2.0$  \hfill \\
  \hline
  Pow &$2.1\times 10^{-3}$&$1.1\times 10^{-3}$&$4.7 \times 10^{-4}$& $2.5\times 10^{-4}$ &$4.4\times 10^{-4}$&$5.9\times 10^{-4}$ \hfill \\
  LI&   $2.0\times 10^{-5}$ & $1.7\times 10^{-5}$&$1.6\times 10^{-5}$ &  $1.5\times 10^{-5}$&$\bf 1.6\times 10^{-5}$&$\bf 1.9\times 10^{-5}$  \hfill \\
  GaSP&  ${ \bf 1.4\times 10^{-5}} $& ${ \bf 1.2\times 10^{-5}}$ & ${ \bf 1.0\times 10^{-5}}$    &${\bf 1.3\times 10^{-5}}$ &${ 1.9\times 10^{-5}}$ &${ 3.6\times 10^{-5} }$ \hfill \\
    QPD&  ${ 1.8\times 10^{-5}}$ &  ${ 1.7\times 10^{-5}}$   &${ \bf 1.0\times 10^{-5}}$&${ 2.6\times 10^{-5}}$&${ 1.1\times 10^{-4}}$&${ 3.4\times 10^{-4}}$ \hfill \\
  \hline

\end{tabular}
\end{center}
   \caption{ $AvgMSE$ for noisy assessments. $U$ is assumed to be exponential for the first three rows and power for the latter three. The number of observation is assumed to be $n=10$ for all case and $\sigma_{\epsilon}=0.005$.}
   \label{tab:randomutility}
\end{table}

Compared to the results of noise-free assessments, all methods have comparatively large MSE for noisy assessments as shown in Table~\ref{tab:randomutility}. Indeed, if the elicited utilities are dominated by  noise, none of the methods work as well as for the noise-free cases. Compared to parametric fitting to the rest of the methods, GaSP still has the smallest MSE in almost all cases. 

\subsubsection{Comparing derivatives}
In \cite{abdellaoui2007loss,abdellaoui2008tractable}, the curvature of utility functions is classified as either concave, convex or of mixed type, using LI or parametric fitting. Such classification is \textit{global} and characterizes the dominant risk attitude implied by the assessed utility function throughout the  entire domain. In the LI method, empirical derivatives of assessed utility points are utilized for estimation of curvature. Let $P_i$ be an observed utility middle point between two neighboring elicited utility points $P_{i-1}$ and $P_{i+1}$. Denote $ S^{-}(P_i)$ and $ S^{+}(P_i)$ as slopes of the straight line between  $P_i$ to $P_{i-1}$ and  $P_i$ to $P_{i+1}$ respectively. $\Delta S(P_i) = S^{+}(P_i) -  S^{-}(P_i)$ is used as the estimate of convexity at point $P_{i}$. The utility function is typically estimated to be concave (convex) if more than $\approx 2/3$  of elicited utility points are estimated to be  concave (convex), otherwise it is denoted as a mixed type.

\begin{figure}[t]
\centering
  \begin{tabular}{cc}
    \includegraphics[scale=.48]{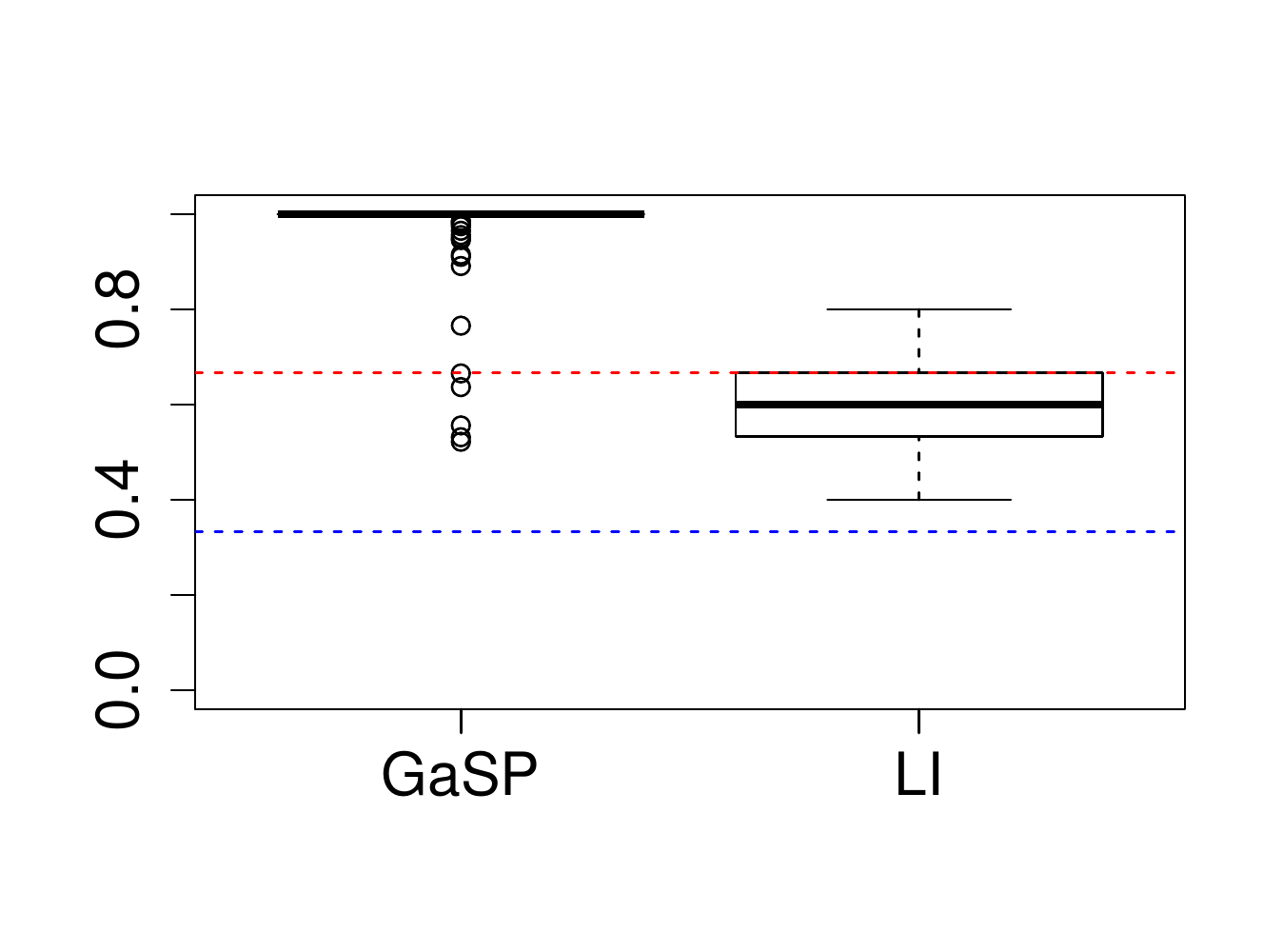}
     \includegraphics[scale=.48]{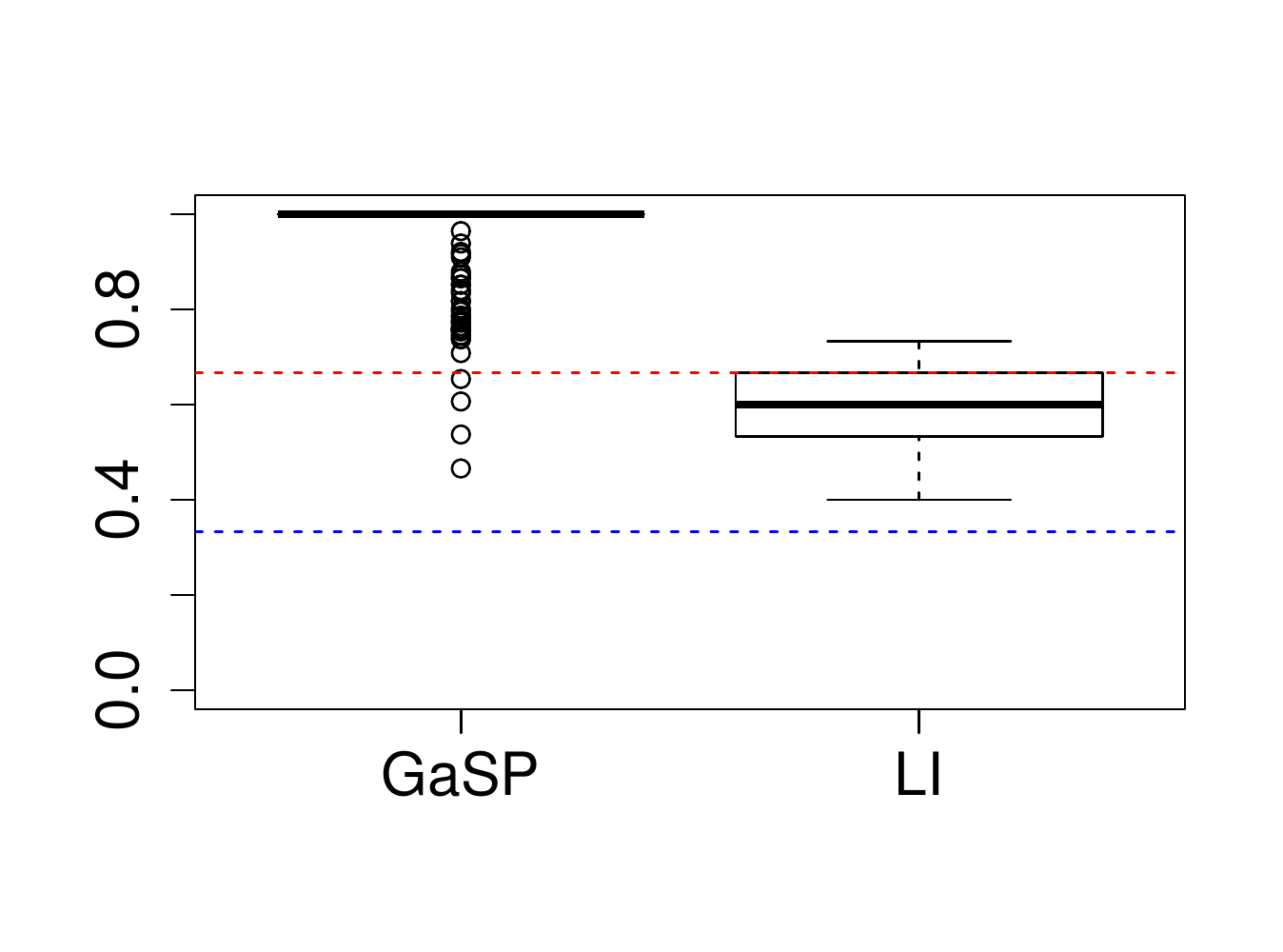}      \vspace{-.5in}\\
         \includegraphics[scale=.48]{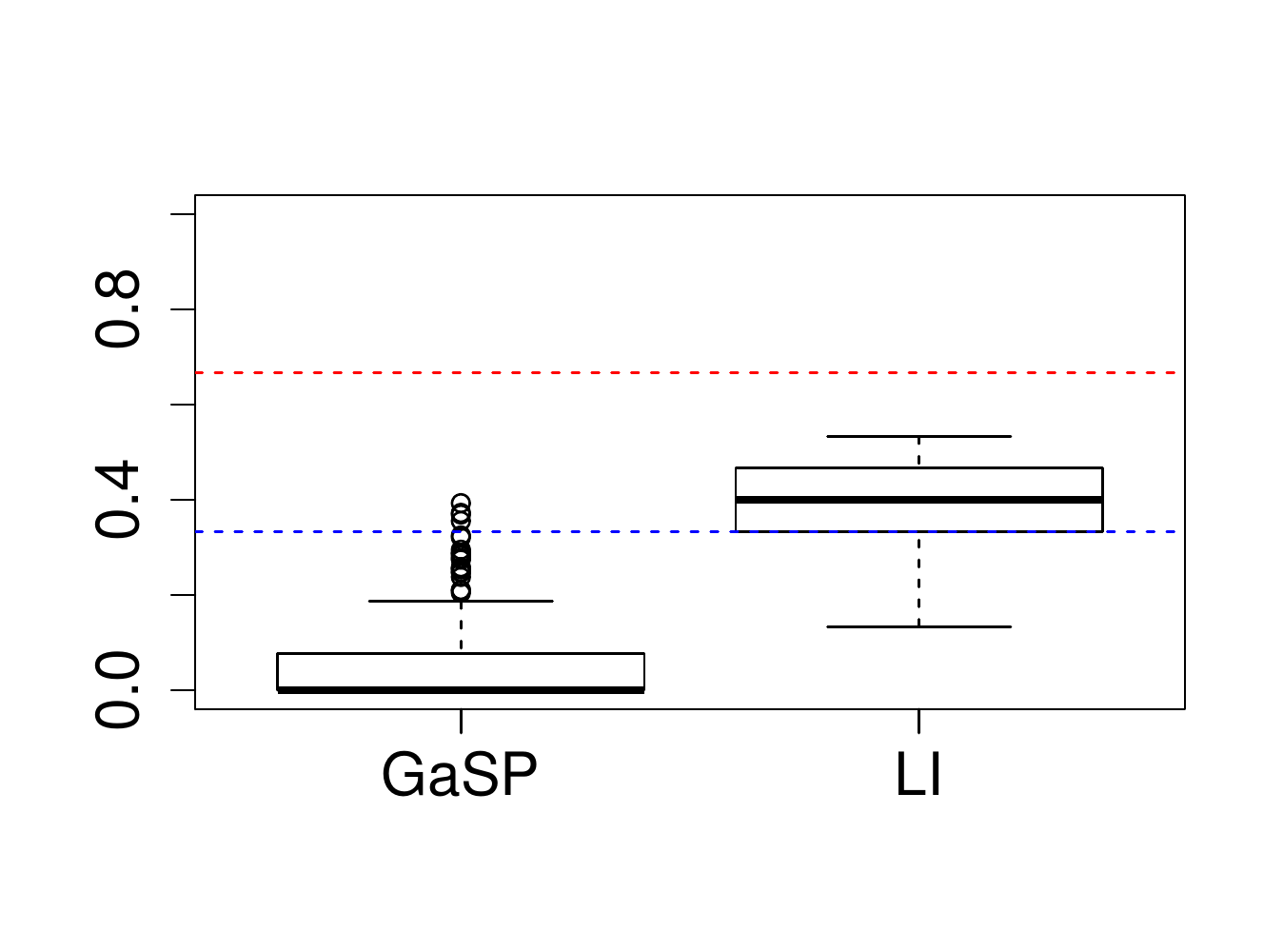}
     \includegraphics[scale=.48]{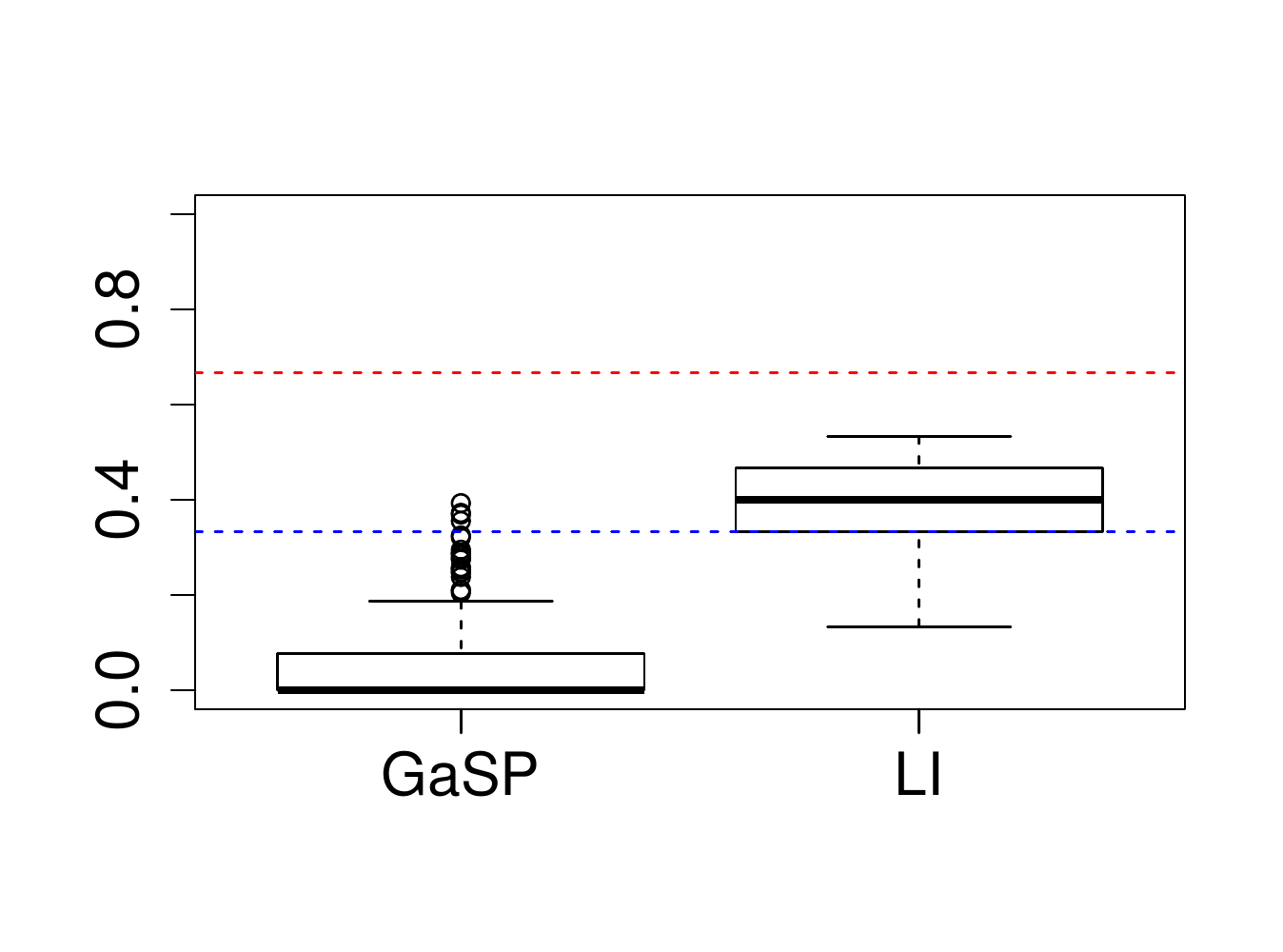}
   \end{tabular}
   \caption{Boxplot of the proportion of points predicted to be {\bf concave} for $N=500$ experiments. Upper figures show results for noisy assessment of concave exponential utilities with $ \sigma_{\epsilon}=0.01, \, \rho=2/3$ (left) and $\sigma_{\epsilon}=0.02,\, \rho=1/3$ (right). Lower figures show results for noisy assessment of convex exponential utilities with $\rho=-1/2,\, \sigma_{\epsilon}=0.01$ (left) and $\rho=-1/4,\, \sigma_{\epsilon}=0.02$ (right).  $n=15$ for all cases. }
\label{fig:global_utility}
\end{figure}

In the GaSP method, one can compute the posterior probability of the second derivatives for \textit{any} point $x^*$ in the domain, $p(u''(x^*) |u(x_1),\ldots,u(x_n))$, as shown in Section~\ref{subsec:model_GP_derivative}. When $p(u''(x^*) \leq 0|u(x_1),\ldots,u(x_n))>0.5$, the utility function is predicted to be concave at $x^*$, otherwise convex. In the following simulations, we consider $n^*=10,000$ equally spaced points and average them out to predict curvature in the GaSP model using the proportion of points that are predicted to be concave/convex.  The results for estimating global concavity by LI and GaSP are shown in Figure~\ref{fig:global_utility}.


In the first row of Figure~\ref{fig:global_utility}, the underlying utility functions are all concave. Due to the effect of noisy assessment, the proportion of concave points by LI is between $1/3$ to $2/3$ in most of the experiments, meaning that LI fails to identify the concavity of the functions, classifying them of mixed type instead. In the second row of Figure~\ref{fig:global_utility}, when the underlying utility function is convex, LI fails to identify the convexity of the utility functions and again classifies a majority of points as the mixed type.

Compared to LI, GaSP predicts the concavity of the functions a lot more accurately. In the upper row, the proportions of concavity points are almost all close to 1 across  $N=500$ experiments.  In the second row, most of the points are predicted to be convex for a majority of experiments.  Using $2/3$ as the threshold would correctly classify most of the utility functions using GaSP.

There are two main reasons why GaSP performs better. First, the GaSP prediction of concavity of point $x^*$ utilizes information from all assessed tuples rather than just the two neighboring points as in LI. Second, the prediction of concavity by GaSP is averaged over many points ($n^*=10,000$ chosen here) rather than the limited number of $n$ assessed tuples in LI.
GaSP estimation is therefore better at analyzing preference properties like risk attitude.

\subsection{Real dataset from \cite{abdellaoui2007loss}}
\label{subsec:realdata}
Let us compare the parametric and nonparametric methods using a real dataset from \cite{abdellaoui2007loss}, collected from a prospect theory based scheme. $k=48$ people answered a series of questions about comparisons between risky gambles. Due to the effect of loss aversion, the range of the loss domain (negative outcomes) is assumed to be $[-1,0]$ and the range of the gain domain (positive outcomes) is constructed to be $[0,0.25]$, with $11$ and $7$ assessed tuples in each domain respectively. The design and elicited scheme are described in  \cite{abdellaoui2007loss}.

\begin{figure}[t]
\centering
  \begin{tabular}{cc}
	\includegraphics[width=.5\textwidth,height=.35\textwidth]{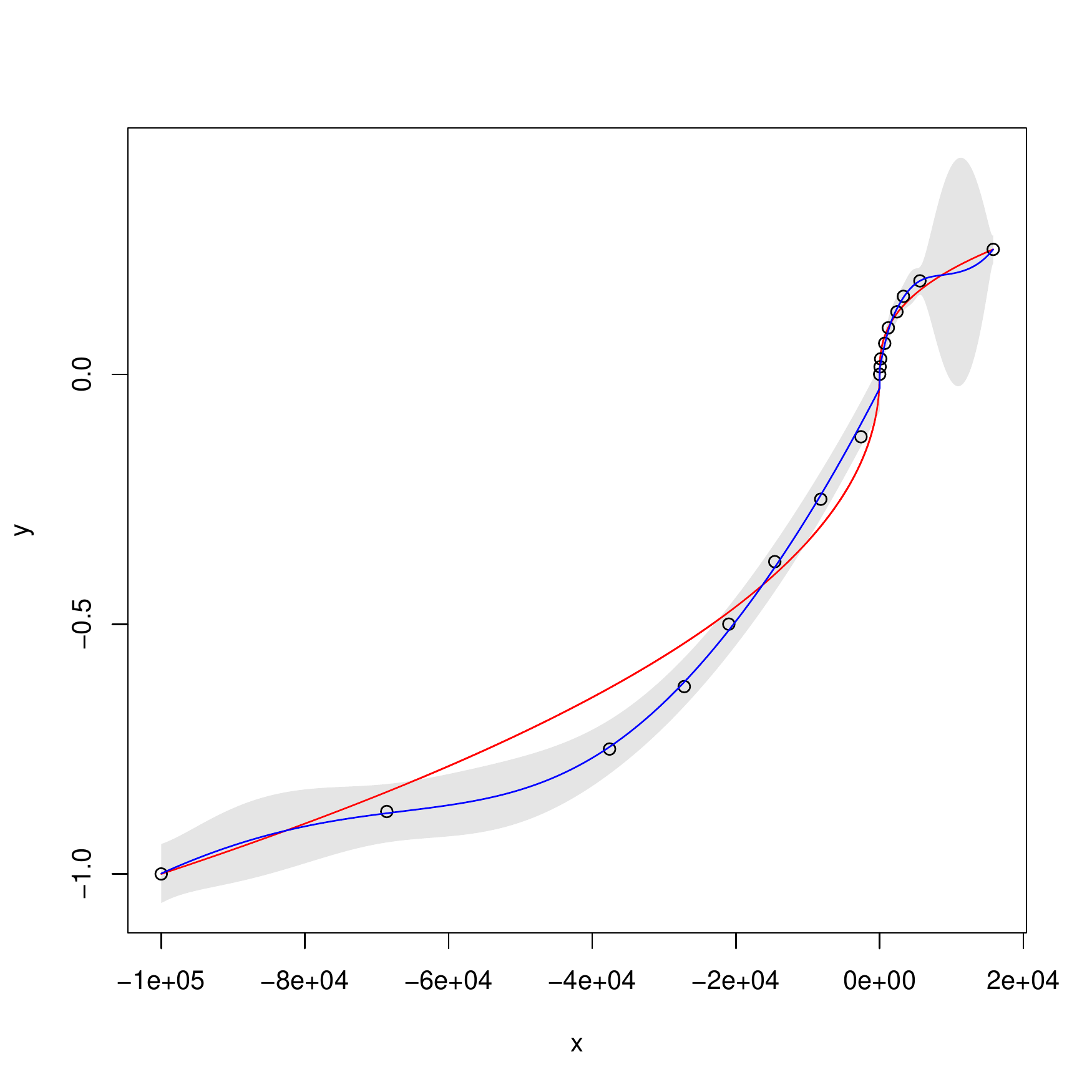}
	\includegraphics[width=.5\textwidth,height=.35\textwidth]{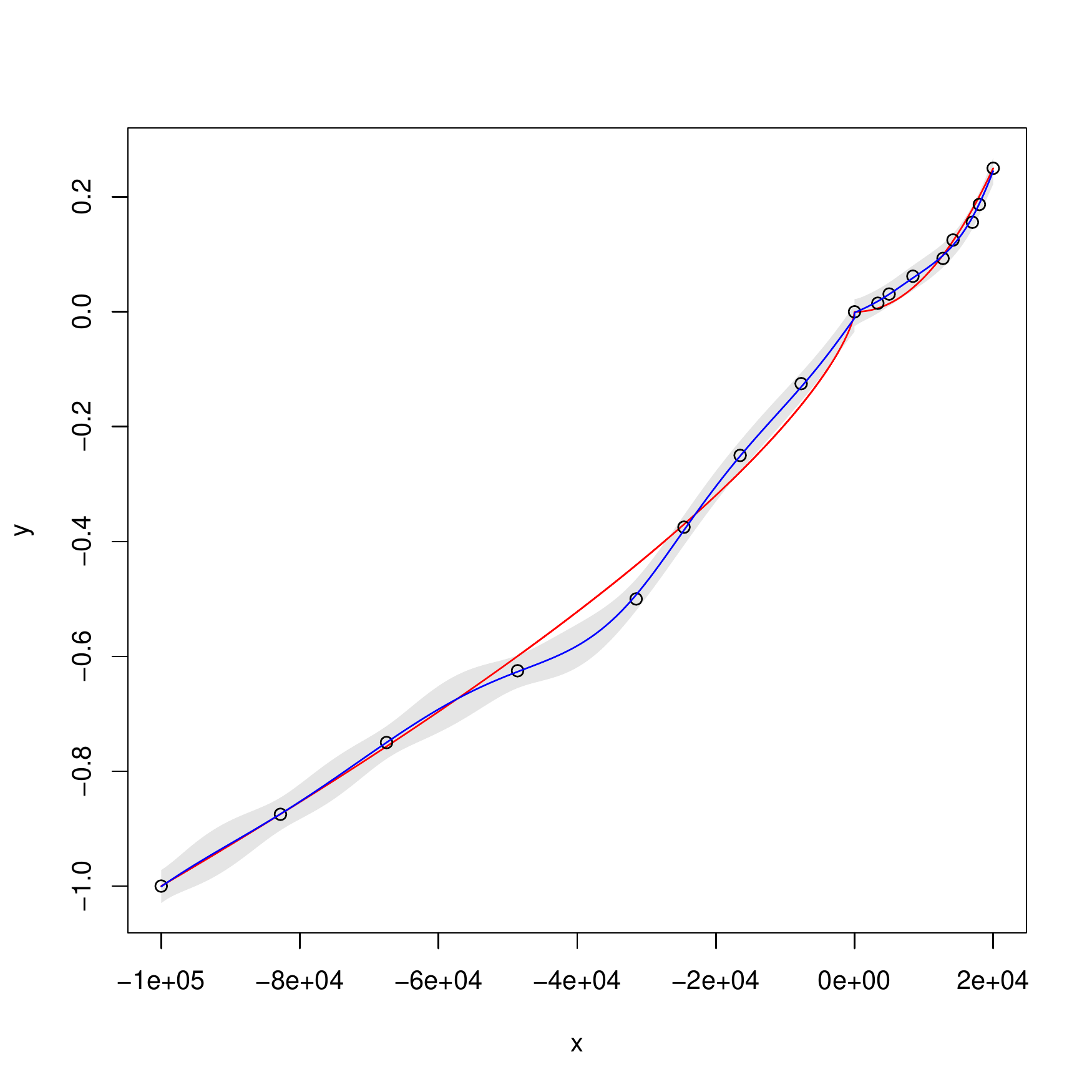}
  \end{tabular}
   \caption{ Utility function estimation for 2 people in \cite{abdellaoui2007loss}. Black circles represent assessed tuples. Blue solid curves are the posterior mean for GaSP and grey shaded area is the 95 \% posterior credible interval; red curves are nonlinear least squares fit for the power function. }
 \label{fig:real1}
\end{figure}

In Figure~\ref{fig:real1}, the assessed tuples and fits for the power function and GaSP along with its 95 \% posterior credible interval are shown for 2 people. Although the power function is widely used in prospect theory (\cite{tversky1992advances,abdellaoui2008tractable}), note that the least squares fit deviates from the assessed curve systematically over a wide range of the domain e.g.  $x \in [-6\times 10^{4}, -2\times 10^{4}]$ for the first person.  For the second person, the power function only seems to fit around $x=-3 \times 10^{4}$. The discrepancy is more likely to be caused by the assumption of the parametric form than the least squares estimation.
In comparison, GaSP seems more flexible for modeling utility functions and fits better in this real dataset.

To test the predictive performance of different methods, we randomly sample $n^*_{loss}=4$ and $n^*_{gain}=3$ assessed tuples in the loss and the gain domain for each person respectively and save them as the test set, while the remaining assessed tuples are used as the training set. We repeat this experiment $N=500$ times, and fit each method at each experiment.

We see in Table~\ref{tab:real1} that the average out of sample MSE for GaSP is similar to that of LI but much smaller than those for the power and exponential functions.  The parametric fit, albeit straightforward in interpretation, induces extra assumptions on the shape of utility functions and may lead to a poor estimate.

\begin{table}[ht]
\begin{center}
\begin{tabular}{lrrrr}
  \hline
    AvgMSE      & GaSP  & LI &Pow&Exp    \\
  \hline
  Loss                   & $8.9\times 10^{-4}$ & $9.7\times 10^{-4}$&$1.5\times 10^{-3}$&$1.7\times 10^{-3}$\\
   Gain             & $7.4\times 10^{-5}$ & $8.2\times 10^{-5}$&$1.1\times 10^{-4}$&$1.1\times 10^{-4}$\\
   \hline

\end{tabular}
\end{center}
   \caption{ Average out of sample MSE for  losses and  gains using GaSP, linear interpolation (LI), power (Pow), exponential (Exp). MSE is averaged over $n_{loss}^*kN=96,000$ and $n_{gain}^*kN=72,000$ respectively.}
   \label{tab:real1}
\end{table}

\section{Multi Attribute Utility Function Estimation}


To demonstrate the performance of the GaSP model for utility functions with multiple attributes, we study a real dataset with three attributes provided in \cite{fischer2000attribute2}. This experiment was conducted among 22 Duke students about decisions pertaining to course selection involving three attributes. The attributes, denoted as $\mathbf x=(x_1,...,x_3)$, include the degree of interest, expected teaching quality and the average grade, each of which has 5 levels. The output, denoted as $u(\mathbf x)$, is the utility rating of a course given a set of attributes.
Each volunteer is asked to rate the same 20 courses along the three attributes.


The \textit{RandMAU} model is proposed in  \cite{fischer2000attribute2} as a sampling model to capture the stimulus properties of \textit{attribute conflict} and \textit{attribute extremity}.
The model is intended to characterize potentially inconsistent responses to  preference elicitation questions through a stochastic model of the assessed utility. It is defined by:
\begin{eqnarray}
      \label{equ:RandMAU}
u(x_1,...,x_p)&=&\sum\limits_{i=1}^p{\omega_iu_i(x_i), } +\omega \sum\limits_{i=1}^p \sum\limits_{j>i}^p \omega_iu_i(x_i)\omega_ju_j(x_j)+...+\omega^{p-1}\prod\limits_{i=1}^p\omega_iu_i(x_i),  \\
      1+\omega&=&\prod\limits_{i=1}^p(1+\omega \omega_i),  \nonumber \\
      \omega_i &\sim_{ind}& Beta(r_i, u_i), \nonumber
\end{eqnarray}
where $p=3$ is the number of attributes and $u_i(x_i)=\left[ \frac{x_i-x_{i0}}{x_i^*-x_{i0}} \right]^{\alpha_i}$ with $x_{i0}$ and $x_i^*$ as the lower and upper limits for attribute  $x_i$. This model has 7 parameters $(\omega_{1}, \omega_{2}, \omega_{3},\omega, \alpha_{1}, \alpha_{2}, \alpha_{3})$ but when $(\omega_{1}, \omega_{2}, \omega_{3})$ are known, $\omega$ can be uniquely solved (as in \cite{fischer2000attribute2}) which leaves the model with 6 degrees of freedom.

\cite{fischer2000attribute2} specify two ways of estimating parameters, namely the \textit{corner point} and \textit{nonlinear least squares} estimation methods. The first approach uses only 7 out of the 20 data points (for each participant) to fit the model while the second minimizes the square error using all assessed tuples.

\begin{table}[ht]
\begin{center}
\begin{tabular}{lrrr}
  \hline
                    &corner point &nonlinear least square & GaSP  \\
  \hline

   $MSE$ &   $0.0084$  & $0.0056$  & $0.0017$     \hfill \\
    $R^2$ &    $0.9087$ &  $0.9391$  & $0.9813$       \hfill \\
   \hline

\end{tabular}
\end{center}
   \caption{Out of sample performance using corner point, nonlinear least square and GaSP estimation.}
   \label{tab:13points_MSE}
\end{table}

\begin{figure}[t]
\centering
  \begin{tabular}{ccc}
    \includegraphics[width=0.33\textwidth]{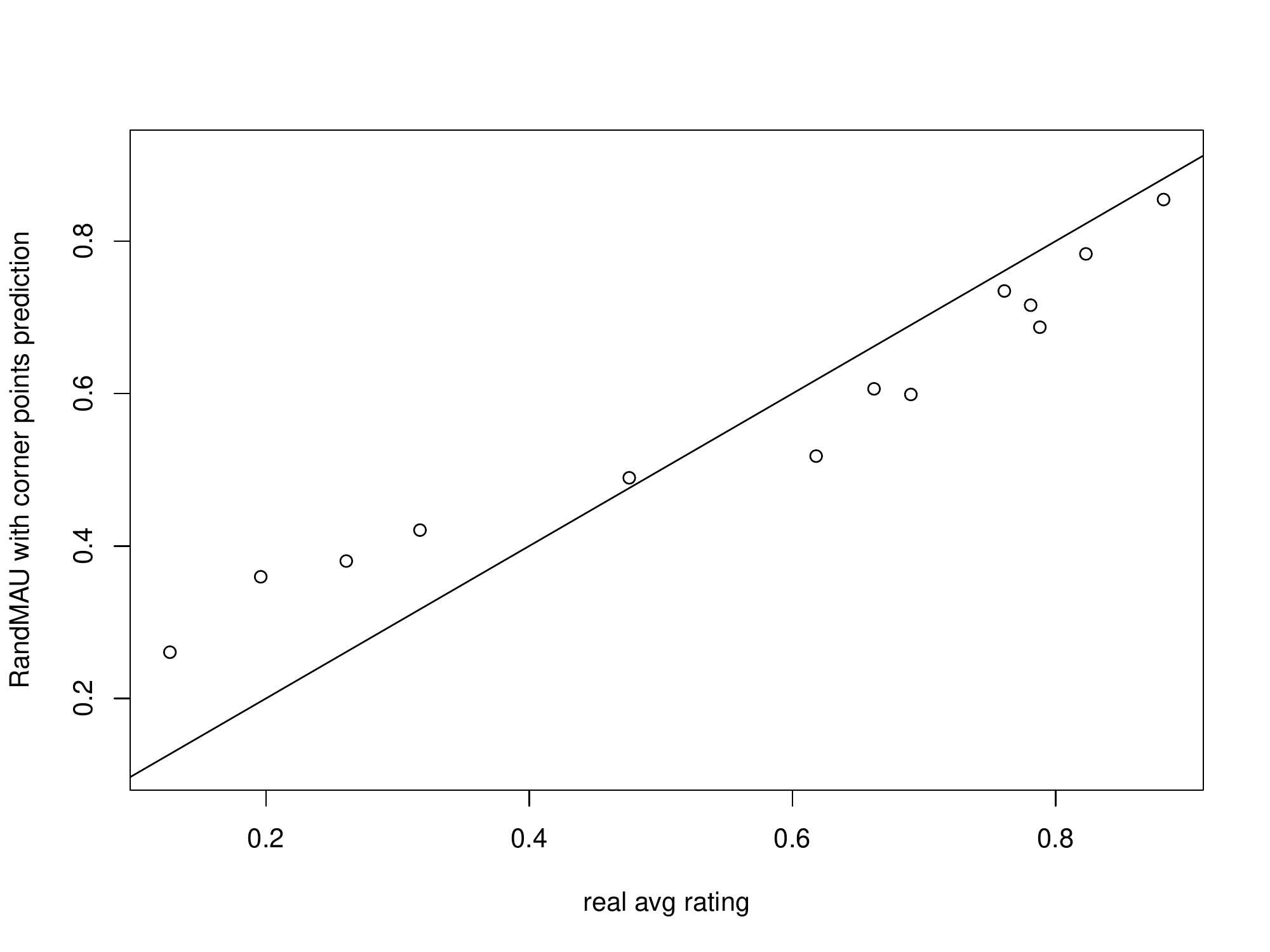}
        \includegraphics[width=0.33\textwidth]{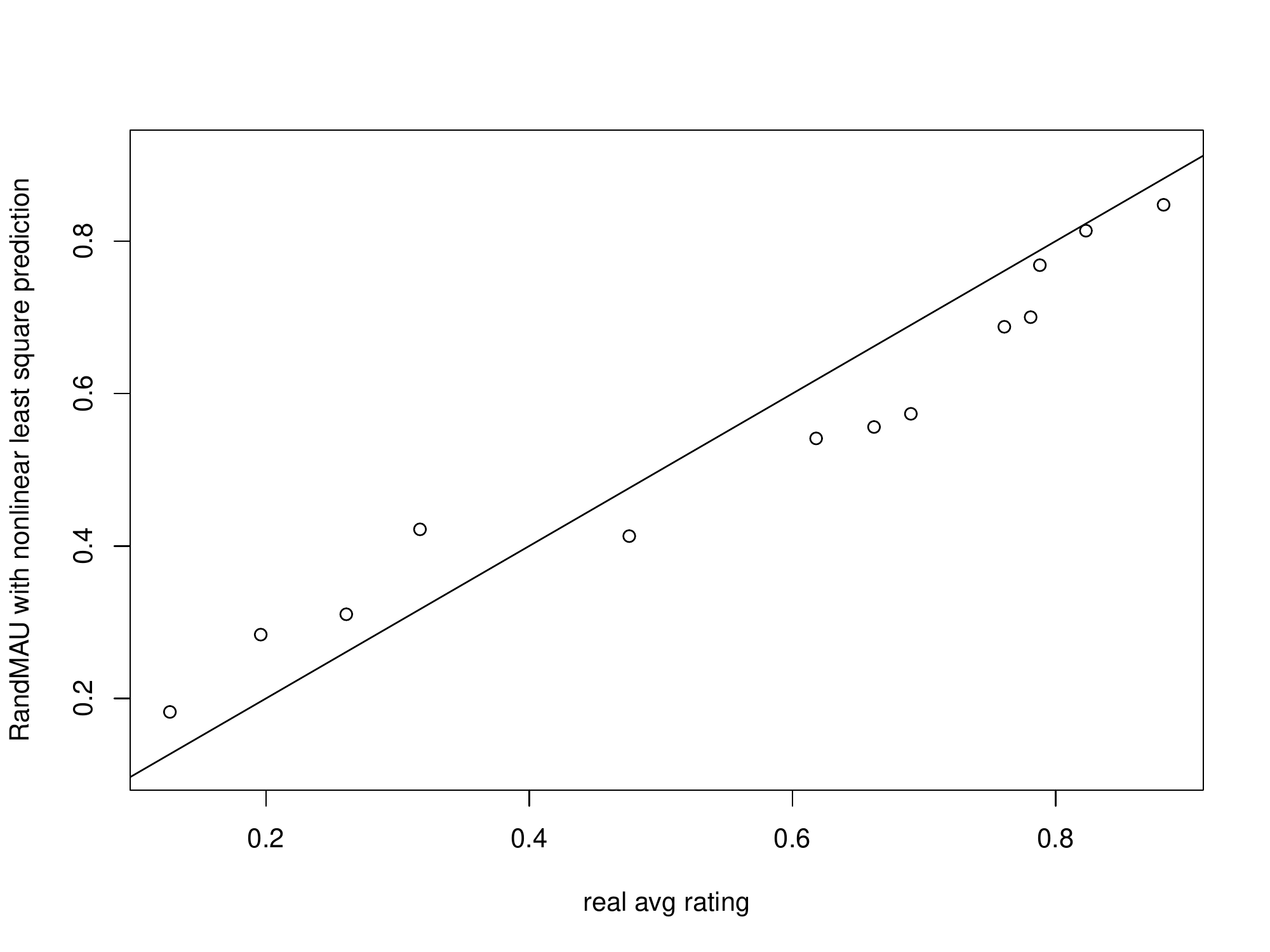}
	\includegraphics[width=0.33\textwidth]{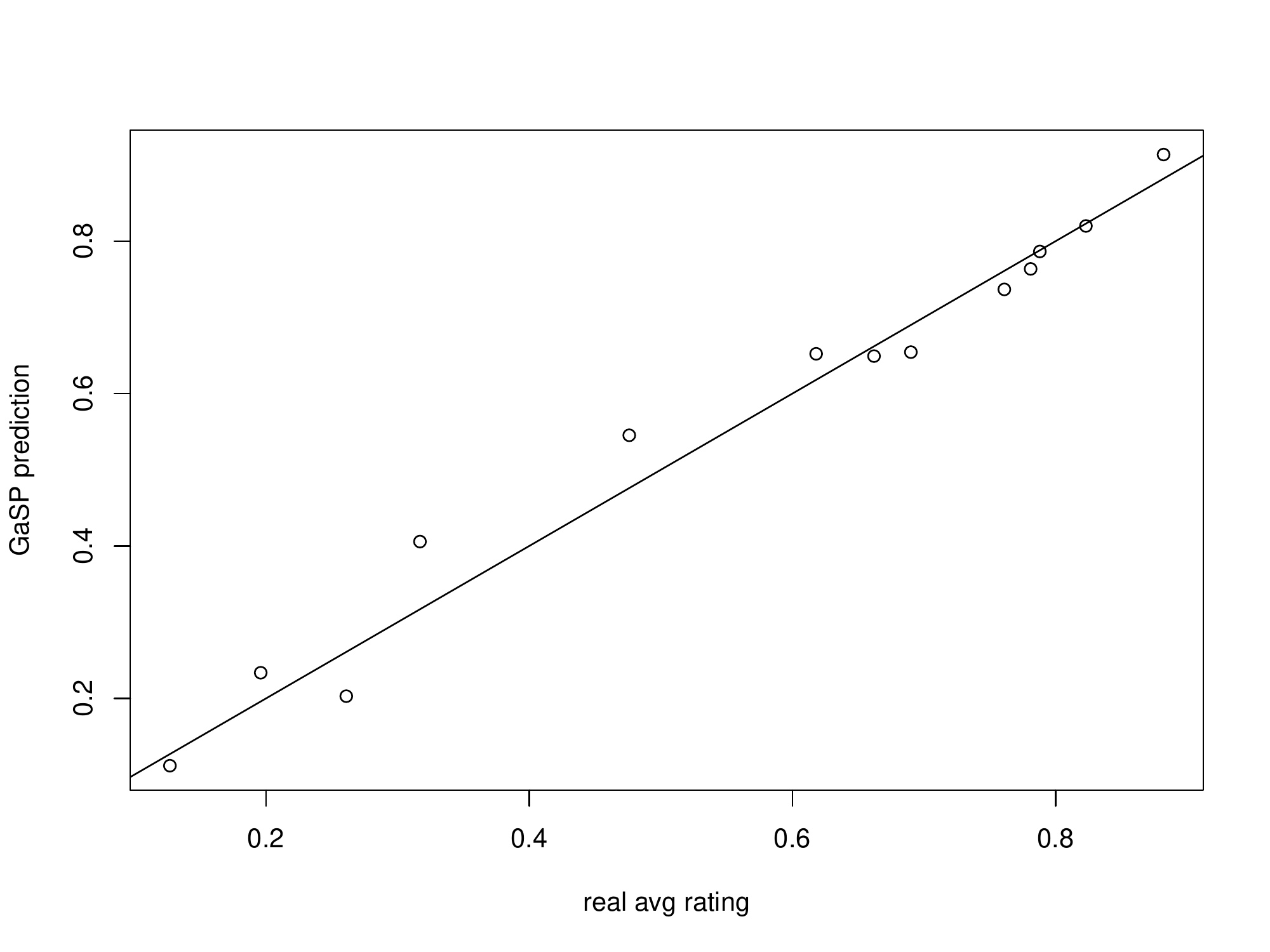}
	\end{tabular}

   \caption{Plot of prediction by RandMAU with corner point estimation (left), nonlinear least square (middle) and GaSP (right) over real values. }
\label{fig:13points_plot}
\end{figure}

We perform experiments to compare the afore-mentioned two approaches with our proposed GaSP approach for  modeling the average ratings (shown in Table 1 in \cite{fischer2000attribute2}) based on the attributes. We compare the out of sample Means Square Error (MSE) and $R^2$ of our proposed GaSP estimation with the RandMAU-based methods of corner point and nonlinear least squares estimation, using the 13 data points that are not used in the corner points approach. Since the sample size is very small, each time we only leave a data point out and compute the MSE and $R^2$ for this point. The average out of sample MSE and $R^2$ are shown in Table~\ref{tab:13points_MSE} and Figure~\ref{fig:13points_plot}.

In Table~\ref{tab:13points_MSE}, we find that the prediction by the GaSP model is several times better than the previous inference methods because it is flexible enough to learn the structure of the utility function through data, while RandMAU is restrictive since each $u_i(x_i)$ is assumed to be the power utility in Equation (\ref{equ:RandMAU}) and thus cannot capture other shapes.  For the RandMAU methods, we find that the nonlinear least square estimates are better than the corner point estimates in terms of MSE and $R^2$. This is because  only 7 observations are used for prediction in corner point estimation, which is statistically inefficient as it simply throws away the rest of the data for fitting the model.

The plot in Figure~\ref{fig:13points_plot} shows the prediction of these 13 average ratings by the different methods vs. the real average ratings. Both corner point estimation and nonlinear least square estimation in RandMAU indicate imbalanced and biased predictions. Small values are all over predicted and larger values are under predicted, perhaps because the form in RandMAU is unable to capture the curvature and latent pattern of the data; the right most figure corresponding to GaSP seems to solve these difficulties.




\section{Conclusions}

We have presented a nonparametric Bayesian approach involving GaSP for inferring a decision maker's utility function using assessed tuples as training data, regardless of the choice of elicitation protocol and underlying theory. We have argued theoretical benefits over parametric approaches around the desired property of interpolation for noise-free assessment. Unlike simpler approaches like linear interpolation, the proposed approach provides for differentiability whenever a differentiable correlation function is chosen.
The Bayesian aspect of our approach provides a probabilistic view of the decision maker's preference properties, such as the utility function and its risk aversion, while the nonparametric aspect provides the flexibility to fit a large class of functions.

Our numerical experiments with simulated, noise-free assessment data show that the GaSP model has lower MSE when compared with parametric fitting from a misspecified function class as well as the quantile parametrized distribution method and linear interpolation.
When the simulated data is corrupted with additive Gaussian noise, GaSP continues to enjoy the least MSE, though all approaches lead to comparatively higher MSE than the noise-free case.

We also demonstrated promising results with real data sets from the literature. These results provide evidence of the effectiveness of using Gaussian processes with respect to out of sample MSE, as well as their flexibility to model empirical utility assessment data.
While the proposed approach is only marginally better than linear interpolation for the single-attribute empirical data set, it provides a principled approach to estimate risk aversion coefficients using the posterior predictive view of the derivative processes. Analysis with the multi-attribute dataset indicates that the proposed estimation method is promising, but further experiments are needed to demonstrate effective estimation in higher dimensions as compared to existing approaches.


There are some unique issues that arise when applying Gaussian processes for estimating utility functions. For instance, a utility function is typically assumed to be monotonic in its arguments.
A potential avenue for future work is around constrained GaSPs, which would extend the proposed approach to respect monotonicity. Even without such considerations, GaSP estimation appears to out-perform the most popular state-of-the-art techniques.

\section*{Appendix}

\proof{Proof of Lemma~\ref{lemma:BM}}

The mean and the covariance of  the Wiener process are $\E[W_t]=0$ for any $t\geq 0$ and $\Cov(W_s, W_t)=min(s,t)$ for any $s, t\geq 0$.  For any $0<t_{i}<t_*<t_{i+1}<0$, the joint distribution of $( W_{t_i} ,  W_{t_*},  W_{t_{i+1}}  )^T$ is
   \[
 \left( {\begin{array}{*{20}{c}}
   W_{t_i} \\
    W_{t_*}  \\
    W_{t_{i+1}} \\
 \end{array} } \right)   \sim {\mathcal{N} } \left( \left(\begin{array}{*{20}{c}}
  {\bf 0} \\
   {\bf 0}  \\
   {\bf 0} \\
 \end{array}
   \right), \,
  \left( {\begin{array}{*{20}{c}}
  t_i & t_i & t_i \\
 t_i  &t_* & t_*\\
t_i & t_* & t_{i+1}\\
 \end{array} } \right) \right),
 \]

By  properties of the multivariate distribution, the conditional distribution of $W_{t_*}| W_{t_i} W_{t_{i+1}}$ is
\[W_{t_*}| W_{t_i} W_{t_{i+1}} \sim \mathcal N(\mu_*, V_*  ), \]
where $\mu_*=\frac{(t_{i+1}-t_*)W_{t_{i}}+ (t_*-t_i)W_{t_{i+1}} }{t_{i+1}-t_i}$ and $V_*=\frac{(t_{i+1}-t_*)(t_*-t_i)}{t_{i+1}-t_i}$. By the Markov property of the process,
\begin{align*}
p(W_{t_*}| W_{t_1},W_{t_2},...,W_{t_n})&=p(W_{t_*}| W_{t_i},W_{t_{i+1}},...,W_{t_n}) \\
&= \frac{p(W_{t_*}, W_{t_i},W_{t_{i+1}},...,W_{t_n})}{p(W_{t_i},W_{t_{i+1}},...,W_{t_n}) } \\
 &= \frac{p(W_{t_*}, W_{t_i},W_{t_{i+1}},...,W_{t_n})}{p(W_{t_{i+2}},W_{t_{i+3}},...,W_{t_{n}}| W_{t_{i}}, W_{t_{i+1}}   ) p(W_{t_{i}},W_{t_{i+1}})} \\
  &= \frac{p(W_{t_*}, W_{t_i},W_{t_{i+1}},...,W_{t_n})}{p(W_{t_{i+2}},W_{t_{i+3}},...,W_{t_{n}}|W_{t_{i}},W_{t_{*}}, W_{t_{i+1}}) p(W_{t_{i}},W_{t_{i+1}})} \\
  &=\frac{p(W_{t_{i}},W_{t_{*}}, W_{t_{i+1}})}{p(W_{t_{i}},W_{t_{i+1}})} \\
  &=p(W_{t_{*}}|W_{t_{i}}, W_{t_{i+1}}).
\end{align*}
The result follows.
\endproof

\proof{Proof of Lemma~\ref{lemma:GaSP_interpolator}}
For any $i$, note that when the prediction is at the design points $ {\mathbf x}_i$
\[\mathbf c^T(\mathbf x^{\mathcal D}_i) \mathbf C^{-1} = \mathbf e^T_i,\]
where  $\mathbf e_i$ is an $n$ dimensional vector with the $i^{th}$ row as $1$ and the others as $0$. Thus
\begin{align*}
\hat{u} (\mathbf x^{\mathcal D}_i )=&  { \mathbf h(\mathbf x^{\mathcal D}_i )} \hat{\bm{\theta}}+\mathbf{c}^T({\mathbf x}^*){{ {\mathbf C}}}^{-1}\left(\mathbf{u}^{\mathcal D}  -{{\mathbf{h}(\mathbf{x}^{\mathcal D}  )}}\hat{\bm{\theta}}\right) \\
=& { \mathbf h(\mathbf x^{\mathcal D}_i )} \hat{\bm{\theta}} +  \mathbf{u}(\mathbf x^{\mathcal D}_i)- \mathbf{h}(\mathbf{x}^{\mathcal D}_i  )\hat{\bm{\theta}} \\
=&\mathbf{u}(\mathbf x^{\mathcal D}_i).
\end{align*}
\endproof
\bibliographystyle{ba}
\bibliography{Reference}


\end{document}